\documentclass[12pt,a4paper]{article}
\usepackage{a4wide}
\usepackage{epsfig}
\usepackage{cite}
\usepackage{amsmath}
\usepackage{graphicx}
\usepackage{latexsym}
\usepackage{array}

\newlength{\absize}
\newcommand{\half}{{\textstyle\frac{1}{2}}}

\def\gsim{\mathrel{\rlap{\raise 2.5pt \hbox{$>$}}\lower 2.5pt
\hbox{$\sim$}}}
\def\lsim{\mathrel{\rlap{\raise 2.5pt \hbox{$<$}}\lower 2.5pt
\hbox{$\sim$}}}

\newcommand{\Lumint}{{\cal L}_{\rm int}}

\usepackage[dvips]{color}
\definecolor{Black}{named}{Black}
\definecolor{Red}{named}{Red}

\begin{document}

\thispagestyle{empty}
\renewcommand{\thefootnote}{\fnsymbol{footnote}}
\newpage\normalsize
\pagestyle{plain}
\setlength{\baselineskip}{4ex}\par
\setcounter{footnote}{0}
\renewcommand{\thefootnote}{\arabic{footnote}}
\newcommand{\preprint}[1]{%
\begin{flushright}
\setlength{\baselineskip}{3ex} #1
\end{flushright}}
\renewcommand{\title}[1]{%
\begin{center}
\LARGE #1
\end{center}\par}
\renewcommand{\author}[1]{%
\vspace{2ex}
{\Large
\begin{center}
\setlength{\baselineskip}{3ex} #1 \par
\end{center}}}
\renewcommand{\thanks}[1]{\footnote{#1}}
\renewcommand{\abstract}[1]{%
\vspace{2ex}
\normalsize
\begin{center}
\centerline{\bf Abstract}\par
\vspace{2ex}
\parbox{\absize}{#1\setlength{\baselineskip}{2.5ex}\par}
\end{center}}

\vspace*{4mm} 

\title{Sneutrino identification in dilepton events at  the LHC} \vfill

\author{P. Osland,$^{a,}$\footnote{E-mail: per.osland@ift.uib.no}
A. A. Pankov,$^{b,}$\footnote{E-mail: pankov@ictp.it} 
N. Paver$^{c,}$\footnote{E-mail: nello.paver@ts.infn.it} and 
A. V. Tsytrinov$^{b,}$\footnote{E-mail: tsytrin@rambler.ru}}

\begin{center}
$^{a}$Department of Physics and Technology, University of Bergen,
Postboks 7803, N-5020  Bergen, Norway\\
$^{b}$The Abdus Salam ICTP Affiliated Centre, Technical
University of Gomel, 246746 Gomel, Belarus\\
$^{c}$University of Trieste and INFN-Trieste Section, 34100
Trieste, Italy
\end{center}
%
%
%
\vfill

\abstract{
Heavy neutral resonances appearing in the clean Drell-Yan
channel may be the first new physics to be observed at the
proton-proton CERN LHC. If a new resonance is discovered at the
LHC as a (narrow) peak in the dilepton invariant mass distribution,
the characterization of its spin and couplings will proceed via the
measurement of production rates and angular distributions of the decay
products. We discuss the discrimination of a spin-0 resonance
(sneutrino) predicted by supersymmetric theories with $R$-parity
breaking against the spin-1 of $Z^\prime$ bosons and the
Randall-Sundrum graviton resonance (spin-2)  with the same mass
and producing the same number of events under the observed peak.
To assess the region of sneutrino parameters
(couplings and masses) where the spin determination can be
performed to a given confidence level, we focus on the event rate
and the angular distributions of the Drell--Yan leptons, in
particular using the center-edge asymmetry, $A_{\rm CE}$. We find
that although the measured event rate permits solving the above problem
partially, the center-edge asymmetry, on the contrary
allows to differentiate the various spins entirely with a minimal
number of events around 200.
}

\section{Introduction} \label{sect:introduction}

One of the first aims of the LHC experiments will be to
search for new high-mass resonances in the inclusive
Drell--Yan production of dilepton pairs
\begin{equation}
p+p\to l^+l^-+X \qquad (l=e,\mu),
\label{proc_DY}
\end{equation}
where the Fermilab experiments have already pushed the mass limit up to
$\sim 1~\text{TeV}$ \cite{Aaltonen:2008ah,Abazov:2010xh}.
If such a resonance should be found, as a peak
in the invariant dilepton mass distribution, it will be
crucial to determine its spin. The default assumption
would be a spin-1 $Z^\prime$, for which
many models have been proposed \cite{Hewett:1988xc}, whereas
the more exotic options would be the Randall-Sundrum (RS)
spin-2 graviton excitation \cite{Randall:1999ee} or
a spin-0 scalar. For
appropriate values of the relevant coupling constants, these models
can lead to the same mass and same number of events under the peak,
hence could not be distinguishable from one another from
statistics alone. The cases of $Z^\prime$ and graviton,
and the perspectives for their discovery and identification, have
been extensively discussed in the literature
\cite{Davoudiasl:2009cd}, and we shall here focus on the third
option, namely a scalar.

The three popular classes of scalars, all hypothetical, are: the
Higgs scalar(s), the gravi\-scalars or radions (associated with
extra dimensions), and those of supersymmetric theories. While the
former two would have a very small branching ratio into a lepton
pair, the (scalar) sneutrino could in an $R$-parity-violating
theory be singly produced, and have a non-negligible branching
ratio into light leptons, $e^+e^-$ and $\mu^+\mu^-$. This is the
case we will study, we will establish in what range
of masses and coupling strengths one can distinguish it from a
$Z^\prime$ and a heavy graviton at the LHC.

Originally, conservation of $R_p=(-1)^{(2S+3B+L)}$,
distinguishing ordinary particles from superpartners, was
assumed in supersymmetric extensions of the Standard Model (SM)
mainly to avoid fast proton decay. However, since such baryon and
lepton-number violating processes can in principle be avoided
(or suppressed) also by some other additional symmetries,
SUSY models where $R$-parity violation can to some extent
occur are viable and the corresponding phenomenological scenarios
have gained considerable interest, see, for example,  
Refs.~\cite{Kalinowski:1997bc,Barbier:2004ez,Bhattacharyya:2010kr}.

For a sneutrino in an $R$-parity-violating theory,
we take the basic couplings to be given by
\begin{equation}
\lambda_{ijk} L_{i}L_{j}{\bar{E}}_{k}
+\lambda_{ijk}^{\prime}L_{i}Q_{j}{\bar{D}_{k}},
\label{Rviol}
\end{equation}
with $i,j,k$ generation indices.
Furthermore, $L$ $(Q)$ are the left-handed lepton (quark)
doublet superfields, and ${\bar{E}}$ (${\bar{D}}$) are the
corresponding left-handed singlet fields. Indeed, due to
$SU(2)$ invariance, the sneutrino, which is a $T_3=+\half$-member
of the doublet, can only couple to a {\it down}-type quark.

The following facts are important:
\begin{itemize}
\item
In Drell-Yan dilepton pair production at $pp$ and $p\bar p$
colliders, the sneutrino can be produced
from a $d\bar d$ annihilation in a color singlet state, neither
from $u\bar u$ nor from initial-state gluons. This is different
from the case of a graviscalar.
\item
There is a wide range of non-excluded coupling strengths,
both to the initial-state $q\bar q$ pair ($\lambda^\prime$) and
to the final-state dilepton pair ($\lambda$). Only the product
of these will be relevant to process (\ref{proc_DY}), 
together with the sneutrino width. In fact, they enter 
via the parameter
\begin{equation}
X=(\lambda_{i11}^\prime)^2\text{BR},
\label{X}
\end{equation}
where BR is the sneutrino leptonic branching ratio
and $\lambda_{i11}^\prime$ the relevant couplings to 
the $d\bar{d}$ quarks, with $i$ denoting the sneutrino 
generation. Among these couplings, $\lambda_{111}^\prime$ is 
rather constrained (by neutrinoless double beta decay), 
whereas $\lambda_{211}^\prime$ and $\lambda_{311}^\prime$ 
could be as large as $10^{-2}$--$10^{-1}$ for sneutrinos 
mass scales in the 100 GeV range \cite{Barbier:2004ez}, and 
perhaps larger for heavier masses. Similar orders of 
magnitude at the grand-unification scale are obtained  
from consideration of neutrino masses 
\cite{Dreiner:2010ye}. Earlier 
estimates can be found, e.g., in~\cite{constraints}.
\end{itemize}

We discuss in this paper the perspectives to
identify the spin-0 sneutrino contribution to
process (\ref{proc_DY}) at the LHC, against the spin-1 $Z^\prime$
boson and the spin-2 RS graviton hypotheses, also taking
into account the current mass limits from the Fermilab
Tevatron. It will be demonstrated that these different
cases can to a large extent be sorted out, from an analysis of
the total cross sections and angular distributions.

Specifically, in Sec.~\ref{sec:cross-sections} we will
for completeness give a minimum of
relevant formulae defining the basic observables used in our
analysis. To make the arguments underlying our discussion more
transparent, a brief account of the angular distributions
characterizing the different models under consideration is given
in Sec.~\ref{sec:parametersanddistributions}.
In Sec.~\ref{sec:confusionandidentification} we
define the concept of ``confusion region'' and discuss
our method of spin identification. In
Sec.~\ref{sec:14TeV-highluminosity} we review the prospects for
discriminating any sneutrino from a $Z^\prime$ or a graviton
candidate, if some signal were to be found in the
14-TeV data, where we assume a generous sample of
$100~\text{fb}^{-1}$. In Sec.~\ref{sec:low-energy}
we discuss the early 7-TeV
data and low luminosity.
Sec.~\ref{Sec:coupling} is devoted to a brief discussion of
the importance of integrated luminosity, as well as
the precision that can be obtained on the sneutrino couplings.

\section{Cross sections and angular observables}
\label{sec:cross-sections}
\setcounter{equation}{0}

\subsection{Event rates and differential cross sections}
The basic experimental observables for discovery of a
resonance peak in reaction (\ref{proc_DY}) and its spin
identification are the total production cross section, governing
the event rate:
\begin{align}
&\sigma{(pp\to R)} \cdot \text{BR}(R \to l^+l^-) \nonumber \\
&=\int_{-z_{\text{cut}}}^{z_\text{cut}}d z \int_{M_{R}-\Delta
M/2}^{M_{R}+\Delta M/2}d M \int_{-Y}^{Y}d y
\frac{d\sigma}{d M\, d y\, d z}, \label{Eq:TotCr}
\end{align}
and the lepton differential angular distribution, allowing
spin discrimination:
\begin{equation}
\frac{d\sigma}{d z} =\int_{M_{R}-\Delta M/2}^{M_{R}+\Delta
M/2}d M \int_{-Y}^{Y}\frac{d\sigma}{d M\, d y\, d
z}\,d y. \label{DiffCr}
\end{equation}
Here, with $M$ the dilepton invariant mass and
$M_R$ the position of the peak: $R$ denotes the three hypotheses for
the nature of the observed peak, namely, $R={\tilde\nu},Z^\prime,G$
for sneutrino, $Z^\prime$ and RS graviton ``$s$-channel'' exchange,
respectively; $z=\cos\theta_{\rm c.m.}$ with $\theta_{\rm c.m.}$
the lepton-quark angle in the dilepton center-of-mass frame and $y$
is the dilepton rapidity. For the full final phase space, the
integration limits in Eqs.~(\ref{Eq:TotCr}) and (\ref{DiffCr})
would be $Y=\log({\sqrt s}/M)$, with $s$ the proton-proton
center-of-mass energy squared, and $z_{\rm cut}=1$. However, to
account for finite angular detector acceptance, $z_{\rm cut}<1$
and $Y$ must be replaced by a maximum value
$y_{\rm max}(z,M)$. Concerning the size of the
dilepton invariant mass bin $\Delta M$, we adopt
the parametrization of $\Delta M$ vs.\ $M$ proposed in
Ref.~\cite{Atlas}. Of course, on the one hand, larger $\Delta M$
could increase the chance of discovery, and, on the other hand,
for an expected narrow peak  falling within the bin the signal
integral over $M$ should be insensitive to the size of $\Delta M$,
whereas the background would increase with $\Delta M$.

To evaluate the event rates and the spin-identification reaches,
we shall in Eqs.~(\ref{Eq:TotCr}) and (\ref{DiffCr}) use the
CTEQ6.5 parton distributions~\cite{Pumplin:2002vw}, and convolute
them with the partonic cross sections relevant to the different
subprocesses under consideration. Also, we shall include 
$K$-factors to account for next-to-leading-order QCD effects.  

Furthermore, we will impose on the final phase space the constraints 
specific to the LHC detectors, namely: pseudorapidity 
$\vert\eta\vert<2.5$ for both leptons assumed massless (this leads to 
a  boost-dependent cut on $z$~\cite{Dvergsnes:2004tw}); lepton transverse 
momentum $p_\perp > 20\, {\rm GeV}$; reconstruction efficiency of 90\% 
for both electrons and muons \cite{Cousins:2004jc}. Finally, denoting
by $N_B$ and $N_S$ the number of ``background'' and ``signal''
events in the $\Delta M$ bin, the criterion $N_S=5{\sqrt{N_B}}$
or 10~events, whichever is larger, will be adopted as the minimum
signal for the peak $R$ discovery.

\subsection{Center-edge asymmetry for spin identification}
As anticipated above, the $z$-dependence of Eq.~(\ref{DiffCr}) is
different for sneutrino, $Z^\prime$ and RS graviton exchanges, so
that in principle they could be discriminated from each other by
performing an angular analysis
\cite{Allanach:2000nr,Cousins:2005pq,Antipin:2009ch,Boudjema:2009fz,Gao:2010qx}.
In practice, this initially requires a boost from the laboratory
frame to the dilepton center-of-mass frame on an event-by-event
basis. In fact, due to the complete symmetry of the $pp$ initial
state at the LHC, the sign of the variable $z$ may not be
unambiguously determined for all measured events. To perform the
angular analysis we use the {\it evenly} integrated angular
center-edge asymmetry, defined as
\cite{Osland:2003fn,Osland:2008sy,Osland:2009tn}:
\begin{align}
\label{ace}
A_{\rm{CE}}&=\frac{\sigma_{\rm{CE}}}{\sigma}\quad
{\rm with} \nonumber \\
\sigma_{\rm{CE}} &\equiv \left[\int_{-z^*}^{z^*} -
\left(\int_{-z_{\rm cut}}^{-z^*} +\int_{z^*}^{z_{\rm cut}}\right)\right] 
\frac{{\rm d} \sigma}{{\rm d} z}\, {\rm d} z.
\end{align}
Here, $0<z^*<z_{\rm cut}$ is a priori free, and defines
the separation between the ``center'' and the ``edge''
angular regions. The actual value of $z^*$ will be
``optimized'' later in the numerical analysis.

The observable $A_{\rm CE}$, requiring symmetric
$z$-integration, should by definition minimize
the ``dilution'' implicit in the determination of
the sign of $z$ mentioned above. Moreover, as shown by the
previous applications of Refs.~\cite{Osland:2008sy} and
\cite{Osland:2009tn} to the identification of RS graviton and
$Z^\prime$ spins respectively, being based on integrated
numbers of events a measurement of $A_{\rm CE}$
could apply to spin identification also
in the presence of limited statistics. Finally, being
defined by ratios of angular-integrated cross sections,
this observable promises to be less sensitive to
systematic uncertainties. This is particularly so for the 
numerical uncertainties related to the $K$-factor values 
input in Eqs.~(\ref{Eq:TotCr}) and (\ref{DiffCr}). $K$-factors 
show some $M$-dependence, mostly in the case of 
gluon-initiated processes through the adopted  
parton distributions and flattening out for increasing dilepton 
masses. Moreover, they depend on the chosen   
definition of the factorization scale vs the 
renormalization one (the most ``popular'' - but not unique -  choice 
being $\mu_F=\mu_R=M_{\tilde\nu}$). These corrections are expected to 
cancel to a large extent from the ratio (\ref{ace}), and this allows 
a rather stable assessment of the minimum number of events 
necessary for spin-identification of the resonance peak by means 
of $A_{\rm CE}$, which is our main purpose here. 
Numerical attempts with different sets of parton distributions 
and $K$-factors essentially confirm this expectation. Conversely, 
the mentioned ambiguities on $K$-factors may directly affect the 
theoretical predictions for total production rates and discovery 
reaches, for example needed for the determination of sneutrino couplings 
from data, and within the partial cancellations found among the different kinds 
of corrections, can amount cumulatively to a 10\% uncertainty.     

We mention, in this regard, that spin-diagnostics for 
$s$-channel produced resonances in the process (\ref{proc_DY}), 
based on the analogous center-edge asymmetry
${\tilde A}_{\rm CE}$ defined in terms of differences
between the lepton and the antilepton pseudorapidities, have 
recently been successfully applied in Ref.~\cite{Diener:2009ee}. 
Also, graviton spin-2 identification from the azimuthal angular dependence 
of graviton+jet inclusive production followed by graviton 
decaying into a dilepton pair has been attempted 
in \cite{Murayama:2009jz}.

\section{Model parameters and angular distributions}
\label{sec:parametersanddistributions}
\setcounter{equation}{0}
As anticipated in the Introduction, the detection of a peak
at some $M=M_R$ may not be sufficient to identify the non-standard
interaction causing it. Indeed, different scenarios
can in certain domains of the relevant coupling constants
give the same number of events under the peak at $M_R$, in
which case the spin identification is crucial. In this regard,
for any model we define as {\it discovery} reach the
upper limit of the resonance mass range $M_R$ for which a signal
can be observed (generally, at 5~$\sigma$), and as
{\it identification} reach the largest $M_R$ for which
the {\it spin} can be identified at a chosen confidence level.
In our case, the spin-0 identification reach on
the sneutrino will be determined by the minimal number of signal events
allowing to exclude both the spin-1 {\it and} the spin-2 hypotheses.
Of course, both the discovery and the identification reaches
depend, among other things, on the energy and luminosity at the LHC.
We now briefly expose the main features of the three models,
in particular we introduce the notations for the coupling constants
relevant to our discussion and their current experimental limits.
\subsection{$R$-parity violating sneutrino and discovery reach}
In this case the relevant leading order quark-initiated process is
$d{\bar d}\to{\tilde \nu}\to l^+l^-$, and for the explicit 
leading-order expression of the cross section we refer to
\cite{Kalinowski:1997bc}. Basically, the cross section depends on
the exchanged sneutrino mass $M_{\tilde\nu}$ and the $R$-parity
breaking parameter $X$ defined in Eq.~(\ref{X}). Present 95\% CL 
lower limits on the sneutrino mass vary from 
$M_{\tilde\nu}>397\,{\rm GeV}$ for $X=10^{-4}$ to
$M_{\tilde\nu}>866\, {\rm GeV}$ for
$X=10^{-2}$ \cite{Aaltonen:2008ah}. The upper value of $X$, of 
the $10^{-2}$ order, reflects the bounds of order $10^{-1}$ on $\lambda_{211}^\prime$ and $\lambda_{311}^\prime$ mentioned in 
the Introduction, while the lower value, $10^{-4}$, is fixed by 
the experimental sensitivity obtainable at the Tevatron.  

\begin{figure}[tbh!] 
\vspace*{-0.cm} \centerline{ \hspace*{-0.0cm}
\includegraphics[width=9cm,angle=0]{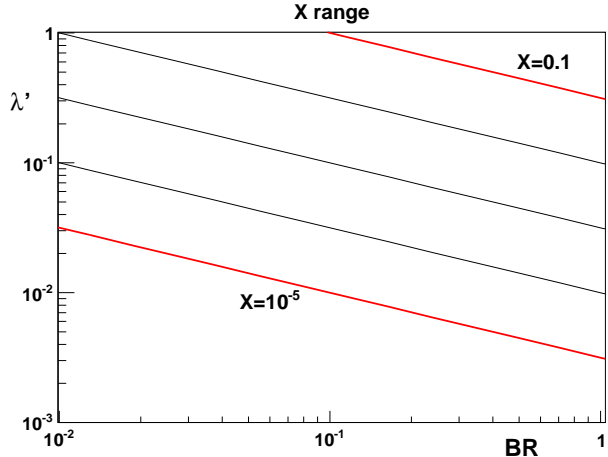}}
\caption{\label{fig-X} Explored range of $X$ vs branching ratio BR. Contours for $X=10^{-5}$, $10^{-4}$, $10^{-3}$, $10^{-2}$ and $0.1$.}
\end{figure}

Current constraints on $X$ (and $\lambda^{\prime}$s) for TeV 
sneutrino masses are very loose, and we will consider the broad 
interval $10^{-5}\leq X\leq 10^{-1}$, which would correspond to 
rather larger values of $\lambda^\prime_{ijk}$, as given by 
Eq.~(\ref{X}) and shown in Fig.~\ref{fig-X} for the range of 
branching values $0.01\leq BR\leq1$. The trilinear leptonic 
couplings of the sneutrino, $\lambda_{ijk}$, are rather model 
dependent, and therefore not specified. On the other hand, 
our main interest here lies in the spin-0 scalar 
exchange identification vs alternative spin exchanges, rather than in 
$\lambda^\prime$ determinations, and in this 
regard our analysis should be considered as 
model-independent. 

The $z$-even dependence of (\ref{DiffCr}) needed in
Eq.~(\ref{ace}) can symbolically be
written in self-explanatory notations as:
\begin{equation}
\frac{{\rm d}\sigma^{\tilde\nu}}{{\rm d} z}
=\frac{3}{8}(1+z^2)\sigma^{\rm SM}_{q}
+ \frac{1}{2}\sigma^{\tilde\nu}_{q}, \label{Diffsneu}
\end{equation}
and at the peak $M=M_{\tilde\nu}$ the spin-0 character
implies a flat angular distribution. The subscript
`$q$' here indicates the $q\bar q$-initiated parton subprocess.
Concerning the center-edge asymmetry:
\begin{equation}
A_{\rm CE}^{\tilde\nu} = \epsilon_q^{\rm SM}\,A_{\rm CE}^{\rm SM}
+\epsilon^{\tilde\nu}_{q}(2z^*-1). \label{acesneu}
\end{equation}
Here, $\epsilon^{\tilde\nu}_{q}$ is the fraction of $q{\bar
q}\to{\tilde\nu}\to l^+l^-$ events, depending on overlaps of
partonic distributions, and $\epsilon_q^{\rm SM}$ is defined
analogously, with $\epsilon^{\tilde\nu}_{q}+\epsilon_q^{\rm SM} = 1$.
Strictly speaking, Eqs.~(\ref{Diffsneu}) and ({\ref{acesneu}) and
the analogous subsequent expressions for the cross sections for the
$Z^\prime$ and graviton cases, hold exactly in the limit
$z_{\rm cut}=1$, whereas we shall impose
$z_{\rm cut}=0.98$. It turns out that the difference is
numerically unimportant at the values of $z^*$ at which the
spin-identification analysis will be performed. The numerical
results presented in the sequel are obtained from ``full''
calculations with all foreseen experimental cuts.

The next-to-leading-order $K$-factors for sneutrino production 
have been evaluated in 
Refs.~\cite{Choudhury:2002aua,Sun:2004gn,Yang:2005ts,ShaoMing:2006dx,Chen:2006ep} 
and, with inclusion of supersymmetric QCD corrections, in 
Ref.~\cite{Dreiner:2006sv}. We can assume that a 
flat value $K=1.30\pm 0.10$ can represent the 
various uncertainties mentioned above, and in Fig.~\ref{fig0} 
we show for this central number the expected 5-$\sigma$ sneutrino 
discovery reaches, for the four cases of LHC center-of-mass energy 
and integrated luminosity anticipated in the Introduction, vs.\
the $R$-parity violating parameter $X$ (the $e^+e^-$ and $\mu^+\mu^-$ 
channels have been combined). In practice, for each line the theoretical 
uncertainties should be encompassed by an approximately 10\% 
uncertainty band. In the remaining part of the paper we will not 
represent those bands in order not to make the plots too busy.   

\begin{figure}[tbh!] 
\vspace*{-0.cm} \centerline{ \hspace*{-0.0cm}
\includegraphics[width=9cm,angle=0]{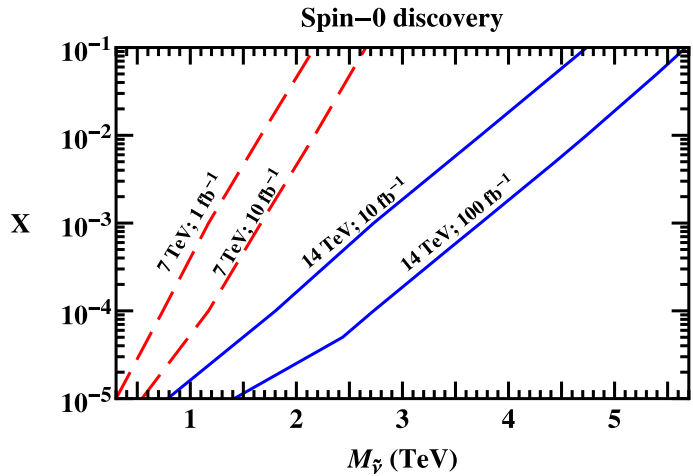}}
\caption{\label{fig0} Discovery reach (5-$\sigma$ level) for
sneutrino production at the LHC in process (\ref{proc_DY}) at
$\sqrt{s}=$7 TeV, $\Lumint=1$ and $10~\text{fb}^{-1}$
(dashed red lines); $\sqrt{s}=$14 TeV, $\Lumint=10$
and $100~\text{fb}^{-1}$ (solid blue lines).}
\end{figure}

\subsection{Extra neutral gauge bosons}
Extra neutral gauge bosons $Z^\prime$ naturally occur in
electroweak models based on extended gauge symmetries, and
the leading-order partonic process
$q{\bar q}\to Z^\prime\to l^+l^-$ should manifest
itself as a narrow peak at $M=M_{Z^\prime}$ with the same
form of the $z$-even differential cross section as for the
SM $\gamma$ and $Z$ exchanges. Thus, the analogues of
Eqs.~(\ref{Diffsneu}) and (\ref{acesneu}) read, using the
same kind of notations:
\begin{equation}
\frac{{\rm d}\sigma^{Z^\prime}}{{\rm d}z}=\frac{3}{8}
(1+z^2)[\sigma_{q}^{\rm SM}+ \sigma^{Z^\prime}_{q}]
\label{Diffzprime}
\end{equation}
and
\begin{equation}
\label{acezprime}
A_{\rm CE}^{Z^\prime}\equiv A_{\rm CE}^{\rm SM}=
\frac{1}{2}z^*(z^{*2}+3)-1.
\end{equation}
Note that, contrary to Eq.~(\ref{acesneu}), in
Eq.~(\ref{acezprime}) there is no $\epsilon$ characterizing
fractions of partonic subprocess  events, and, consequently,
the value of $A_{\rm CE}$ is for {\it all} $Z^\prime$ models the
same as for the SM. Again, a $K$-factor value of 1.30 will be 
assumed for all $Z^\prime$ models \cite{Carena:2004xs}.

We will here consider the following popular scenarios:
$Z^\prime_\chi$, $Z^\prime_\psi$, $Z^\prime_\eta$,
$Z^\prime_{\rm LR}$, $Z^\prime_{\rm ALR}$ models, and the
``sequential'' $Z^\prime_{\rm SSM}$ model with the same
couplings as the SM. Although not fully exhaustive of the
rather numerous multitude of $Z^\prime$ models, the considered
list should nevertheless be sufficiently representative of
the different extended gauge electroweak symmetries
underlying such scenarios. A common feature of the chosen models
is that the $Z^\prime$ couplings to leptons and quarks
are there fixed to well-defined numerical values.
As will be seen below, in Figs.~\ref{fig3-N-14-100}--\ref{fig6-N-7-1},
the considered range of couplings is for $Z^\prime$ models much
more constrained than for the sneutrino and graviton models.
This is caused by assumptions of some kind of unification at high
energies. Detailed descriptions and the relevant values of the
vector and axial-vector $Z^\prime$ couplings needed for the
calculation of the cross sections can be found, e.g., in the
reviews~\cite{Hewett:1988xc}. For the chosen $Z^\prime$ models, 
we list for convenience in Table~I  the values of the left- and right-handed 
couplings used in our later plots, and their combinations relevant to the 
forward-backward-symmetric integration in Eq.~(\ref{ace}).

\begin{table}[htb]
\caption{\label{Table:couplings} Left- and right-handed
couplings of SM fermions to the $Z^\prime$
gauge bosons \cite{Osland:2009tn}.}
\begin{center}
\begin{tabular}{|c|c|c|c|c|c|c|c|c|c|} \hline
fermion  & \multicolumn{3}{|c|}{$e$} & \multicolumn{3}{|c|}{$u$} & \multicolumn{3}{|c|}{$d$} \\
\hline 
  & $g_{\text{L}}$ & $g_{\text{R}}$ & $g_{\text{L}}^2+g_{\text{R}}^2$ 
  & $g_{\text{L}}$ & $g_{\text{R}}$ & $g_{\text{L}}^2+g_{\text{R}}^2$ 
  & $g_{\text{L}}$ & $g_{\text{R}}$ & $g_{\text{L}}^2+g_{\text{R}}^2$ \\
\hline
$Z^\prime_\chi$ & $0.70$ & $0.23$ & $0.54$ & $-0.23$ & $0.23$ & $0.11$& $-0.23$ & $-0.70$ & $0.54$\\
\hline
$Z^\prime_\psi$ & $0.30$ & $-0.30$ & $0.18$ & $0.30$ & $-0.30$ & $0.18$ & $0.30$ & $-0.30$ & $0.18$\\
\hline 
$Z^\prime_\eta$ & $0.19$ & $0.38$ & $0.18$ & $-0.38$ & $0.38$ & $0.29$ & $-0.38$ & $-0.19$ & $0.18$\\
\hline
$Z^\prime_\text{LR}$ & $0.40$ & $-0.40$ & $0.33$ & $-0.13$ & $0.67$ & $0.47$ & $-0.13$ & $-0.94$ & $0.90$\\
\hline
$Z^\prime_\text{ALR}$ & $-0.08$ & $-0.05$ & $0.009$ & $-0.01$ & $0.07$ & $0.005$ & $-0.01$ & $-0.02$ & $0.001$\\
\hline
$Z^\prime_\text{SSM}$ & $-0.64$ & $0.55$ & $0.71$ & $0.82$ & $-0.37$ & $0.81$ & $-1.00$ & $0.18$ & $1.04$\\
\hline
\end{tabular}
\end{center}
\end{table}

Current experimental lower limits on $M_{Z^\prime}$, from
the Tevatron collider \cite{Aaltonen:2008ah}, are model dependent
and range from 878 GeV for $Z^\prime_\psi$ to 1.03 TeV
for $Z^\prime_{\rm SSM}$ (95\% CL). Somewhat higher limits may in
principle be obtained in some cases from electroweak precision
data, see, e.g., Ref.~\cite{Erler:2009jh}. Also, attempts
to identify $Z^\prime$ couplings, once the spin-1 is established,
have been proposed, see for instance
Refs.~\cite{Dittmar:2003ir,Feldman:2006wb,Petriello:2008zr,Salvioni:2009mt}.

\subsection{Graviton excitation from small warped extra dimensions}
The simplest version of RS models \cite{Randall:1999ee}, that
addresses the so-called gauge hierarchy problem $M_{\rm Pl}\gg
M_{\rm EW}$, consists of one warped extra spatial dimension and
two three-dimensional branes at a compactification distance $R_c$,
such that the ordinary SM fields are localized on one brane and
gravity originates from the other one and can propagate in the
full 5-dimensional space. Due to its peculiar geometry, the model
predicts a gravity effective mass scale on the SM brane
$\Lambda_\pi={\bar M}_{\rm Pl}\exp{(-k\pi R_c)}$, where $k$ is the
5-dimensional curvature and ${\bar M}_{\rm Pl}$ the ``reduced''
Planck mass ${\bar M}_{\rm Pl}=1/\sqrt{8\pi G_{\rm N}}$. For
$kR_c\simeq 12$, the parameter $\Lambda_\pi$ is of the TeV order, hence in the
reach of the LHC. The model also predicts a tower of spin-2 graviton
excitations, with masses and (universal) coupling constant to SM
particles of order $\Lambda_\pi$ and $1/\Lambda_\pi$,
respectively, that can be exchanged and searched for in the process
(\ref{proc_DY}) as narrow peaks in $M$. The phenomenologically
most convenient parameterization of the model is in terms of
$M_G$, the mass of the lightest graviton excitation $G$, and the
dimensionless ratio $c=k/{\bar M}_{\rm Pl}$. The theoretically
``natural'' ranges for the above parameters are $0.01\leq c\leq
0.1$ and $\Lambda_\pi< 10$ TeV \cite{Davoudiasl:2000jd}.

Recent experimental limits (95\% CL) from the Fermilab Tevatron
($e^+e^-$ channel) range from $M_G>600$ GeV for $c\cong 0.01$ up
to $M_G>1.05$ TeV for $c\cong 0.1$ \cite{Abazov:2010xh}.

The cross sections for the leading-order subprocesses $q{\bar
q}\to G\to l^+l^-$ and $gg\to G\to l^+l^-$ read
\cite{Han:1998sg,Giudice:1998ck}:
\begin{equation}
\frac{{\rm d}\sigma^G}{{\rm d}z}=
\frac{3}{8}(1+z^2)\sigma_{q}^{\rm SM} +
\frac{5}{8}(1-3z^2+4z^4)\sigma^G_{q} +
\frac{5}{8}(1-z^4)\sigma^G_{g}, \label{Diffg}
\end{equation}
and the center-edge asymmetry
\begin{align}
A_{\rm CE}^{G} 
&=\epsilon_q^{\rm SM}\,A_{\rm CE}^{\rm SM} +
\epsilon^G_q\left[2\,{z^*}^5+\frac{5}{2}\,z^*(1-{z^*}^2)-1\right]
\nonumber \\
&+ \epsilon^G_g\left[\frac{1}{2}\,{z^*}(5-{z^*}^4)-1\right].
\label{aceg}
\end{align}
Analogous to Eq.~(\ref{acesneu}), $\epsilon^G_q$, $\epsilon^G_g$
and $\epsilon_q^{\rm SM}$ in Eq.~(\ref{aceg}) are the fractions
of resonant $G$-events for $q\bar q,gg\to G\to l^+l^-$ and SM
background, respectively, with
$\epsilon^G_q+\epsilon^G_g+\epsilon_q^{\rm SM}=1$.
Next-to-leading-order QCD corrections and their uncertainties have been 
calculated and thoroughly discussed in Ref.~\cite{Mathews:2005bw}. For 
the subsequent calculations we can take a flat $K$-factor determination  $K=1.30$, 
the uncertainty on the total graviton production cross section 
would be comparable to that for the sneutrino. Once more, the impact 
of the $K$-factor on the center-edge asymmetry $A_{\rm CE}$ would be small. 
\section{Confusion regions and spin identification}
\label{sec:confusionandidentification}
\setcounter{equation}{0}
As alluded to in the Introduction, the predicted number
of events under a peak at $M=M_R$, $N_S$, can be the same for
the different spin hypotheses for values of the respective
parameters in specific sets within the domains allowed by
current experimental (and/or theoretically natural) limits.
For any model, one can define
a corresponding {\it signature space} as the region in
the ($M_R,N_S$) plane that can be ``populated'' by varying
its parameters in the above-mentioned allowed domains. Of
course, such domains depend on the LHC energy and
integrated luminosity. On the other hand, the {\it confusion
region} for each pair of models is defined as the domain
determined by the intersection of the respective signature 
spaces, where they can give the same number of events and
therefore cannot be distinguished from each other by event 
rates only.

Basically, the procedure we shall use for model identification,
specifically the sneutrino identification, goes along the following
lines. We assume that a peak is discovered at some value $M=M_R$,
and make the further assumption that it is consistent
with the spin-0 sneutrino hypothesis, i.e., $R=\tilde\nu$. The
assessment of the domain in ($M_{\tilde\nu},X$) where one can
exclude the competitor hypotheses, spin-1 and spin-2 with same
number of peak events at $M_{\tilde\nu}$ and therefore confirm
the spin-0 assumption, starts from the deviations
\begin{equation}
\Delta A_{\rm CE}^{G-{\tilde\nu}}=A^G_{\rm CE}-A^{\tilde\nu}_{\rm CE}
\qquad{\rm and}
\qquad
\Delta A_{\rm CE}^{{Z^\prime}-{\tilde\nu}}=A^{Z^\prime}_{\rm CE}
-A^{\tilde\nu}_{\rm CE}.
\label{deltaGSV}
\end{equation}
Figure~\ref{fig2} shows the example of sneutrino exchange with
$M_{\tilde\nu}=3$ TeV and $X=4\times 10^{-3}$, tested against
spin-1 and spin-2 exchanges with the same number of events, at LHC
with $\sqrt s=14$ TeV and $\Lumint=100$ fb$^{-1}$ [$l=e,\mu$
combined], vertical bars are 1-$\sigma$ statistical uncertainties.
The figure shows that $A_{\rm CE}$ at $z^*\simeq 0.5$ indeed
discriminates the three spin hypotheses at the assumed luminosity,
and indicates that the maximum sensitivity to sneutrino can be
obtained at this particular value of $z^*$. Actually, this property
persists for all cases considered here, therefore in the remaining
part of the paper we shall perform our $A_{\rm CE}$ analysis and
present the corresponding results at $z^*=0.5$.

\begin{figure}[tbh] 
\centerline{ \hspace*{-0.0cm}
\includegraphics[width=7.5cm,angle=0]{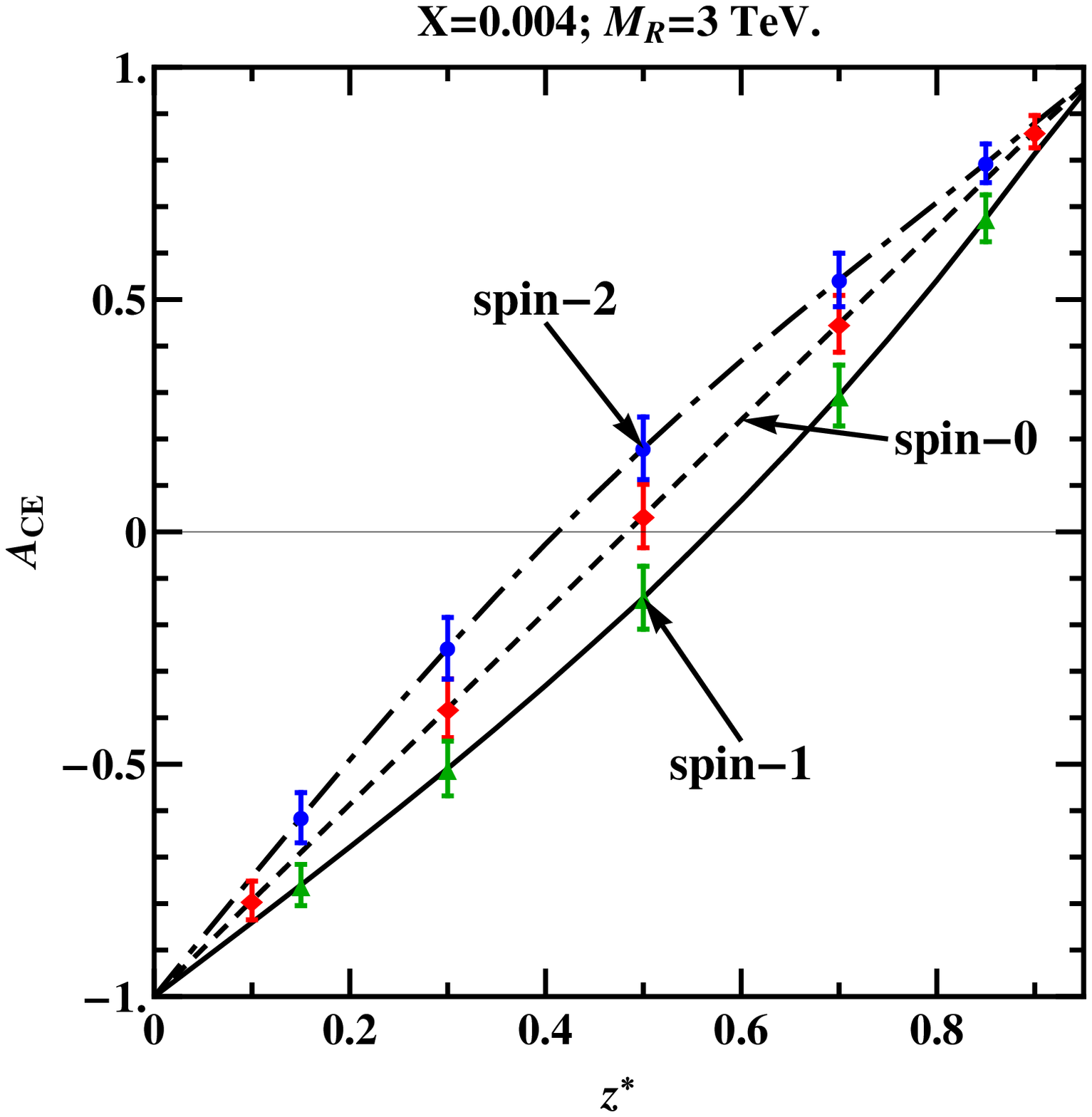}
\hspace*{0.4cm}
\includegraphics[width=7.5cm,angle=0]{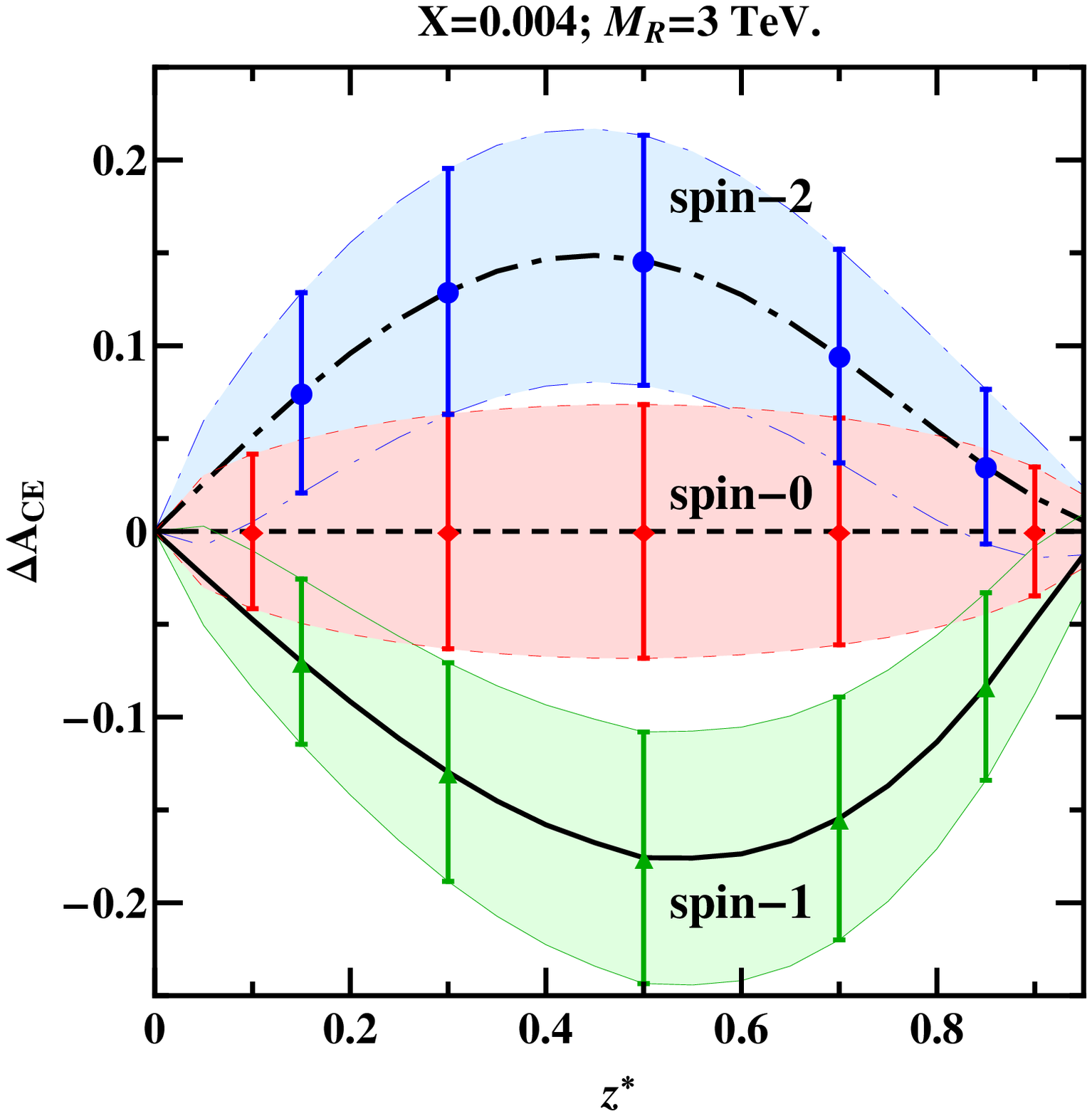}}
\caption{\label{fig2}Left panel: $A_{\rm {CE}}$ vs.\ $z^{*}$
for various hypotheses: $\tilde{\nu}$ (spin-0), $Z'$ (spin-1)
and $G$ (spin-2) with equal mass of 3 TeV. The error bars
attached to the dashed curve are the 1-$\sigma$ statistical
uncertainties on $A_{\rm CE}^{\tilde\nu}$ for the sneutrino with
$X=4\cdot 10^{-3}$ at $100~\text{fb}^{-1}$. The error bars
attached to the solid and dash-dotted curves refer to the
spin-1 and spin-2 hypotheses, respectively, assuming the
same number of resonance events as for the sneutrino case.
Right panel: corresponding $A_{\rm CE}$ deviations
of Eq.~(\ref{deltaGSV}).}
\end{figure}

\par
To define an ``estimator'' for the minimal number of events
needed to exclude the spin-1 and spin-2 hypotheses, we compare the
deviations (\ref{deltaGSV}) with the statistical uncertainty
on $A^{\tilde\nu}_{\rm CE}$ (systematic uncertainties can
also be accounted for in this analysis):
\begin{equation}
\delta A^{\tilde\nu}_{\rm CE}
= \sqrt{\frac{1-(A^{\tilde\nu}_{\rm CE})^2}{N_S}}\approx
\sqrt{\frac{1}{N_S}}.
\label{delace}
\end{equation}
 (Numerically,
$(A^{\tilde\nu}_{\rm CE})^2\ll 1$ at $z^*\simeq 0.5$.)
We apply simple-minded $\chi^2$-like conditions to determine
$N_{\rm min}$ for exclusion of the two competing hypotheses,
respectively:
\begin{equation}
\vert\Delta A_{\rm CE}^{G-{\tilde\nu}}\vert^2=\kappa
\vert\delta A^{\tilde\nu}_{\rm CE}\vert^2; \qquad
\vert\Delta A_{\rm CE}^{{Z^\prime}-{\tilde\nu}}\vert^2 =\kappa
\vert\delta A^{\tilde\nu}_{\rm CE}\vert^2.
\label{chisquare}
\end{equation}
Here, $\kappa$ is a critical value that defines the confidence level,
we take $\kappa=3.84$ for 95\% CL. Also, $N_{\rm min}$ can be
different in order to exclude the spin-1
or the spin-2 hypothesis from the spin-0 assumption, the
larger minimal number of events will be the one needed for
the exclusion of {\it both} competitor
scenarios. In Eq.~(\ref{chisquare}),
$\Delta A_{\rm CE}^{G-{\tilde\nu}}$ and
$\Delta A_{\rm CE}^{{Z^\prime}-{\tilde\nu}}$ of
Eq.~(\ref{deltaGSV}) can be expressed in terms of the resonance
mass $M_R$ and coupling constants ($X$, $c$) with the help
of Eqs.~(\ref{acesneu}), (\ref{acezprime}) and (\ref{aceg}).
Then, the condition (\ref{chisquare}) will define the domains
in the confusion-region planes of model parameters defined above,
where the spin-0 hypothesis can be discriminated from the
others by using the observable $A_{\rm CE}$.
In the next sections we show the signature spaces
and confusion regions relevant to the
LHC running conditions anticipated in the Introduction,
taking into account the current experimental and ``theoretical''
limits exposed above.
\section{LHC nominal energy, high luminosity}
\label{sec:14TeV-highluminosity}
\setcounter{equation}{0}
Starting from the case $\sqrt s=14$ TeV, and high luminosity
${\cal L}_{\rm int}=100\,\text{fb}^{-1}$, the sneutrino signature
space in ($M_R,N_S$) is represented by the full area bounded by
the solid curves labelled as $X=10^{-5}$ and $X=10^{-1}$ in
Fig.~\ref{fig3-N-14-100}, where the relevant phase space cuts
specified above have been applied and the long-dashed line
represents the minimum signal for resonance discovery at $5\sigma$
above the SM ``background''.

\begin{figure}[tbh!] 
\vspace*{-0.5cm} 
\centerline{ 
\hspace*{-1.0cm}
\includegraphics[width=9cm,angle=0]{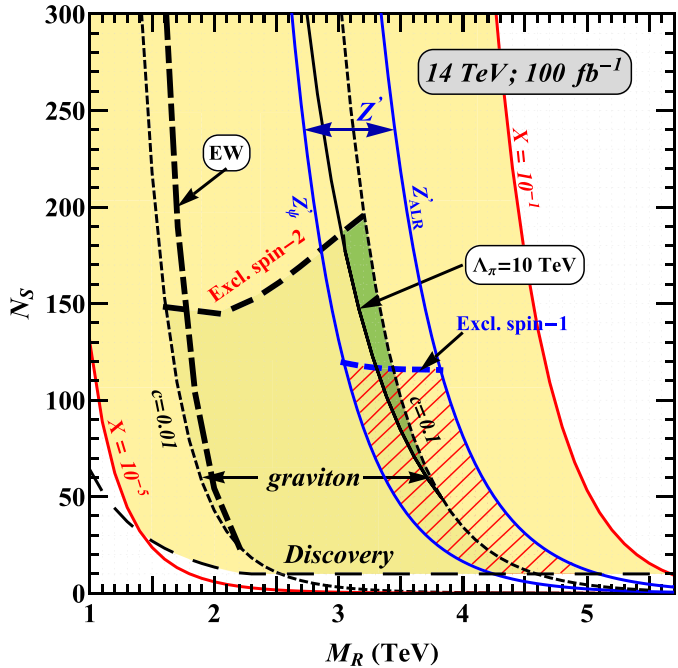}}
\caption{\label{fig3-N-14-100} Number of resonance (signal) events
$N_S$ vs.\ $M_{R}$ ($R=\tilde\nu_\tau,Z^\prime,G,$) at the LHC
with $\Lumint=100~\text{fb}^{-1}$ for the process $pp\to R\to
l^+l^-+X$ ($l=e,\mu$). Yellow area corresponds to the sneutrino
signature space for $10^{-5}<X<10^{-1}$.
The graviton signature space for $0.01<c<0.1$ and
event rates for various $Z^\prime$ models are also shown. Minimum
number of signal events needed to detect the resonance above the
background at the 5-$\sigma$ level (long-dashed curve) and the minimum
number of events to exclude the spin-2 and spin-1 hypotheses in
favor of the spin-0 one, assuming spin-0 is true, at 95\% C.L., are
shown. Regions of confusion are indicated: for sneutrino
and graviton (green); for sneutrino and $Z'$ bosons (hatched).}
\end{figure}

In turn, the signature space for the
lightest graviton RS excitation $G$ is represented by the domain
constrained by the curves ``$c=0.01$'' and ``$c=0.1$''. However, the
portion lying to the left of the curve labelled as
$\Lambda_\pi=10$ TeV is theoretically disfavored. The curve ``EW'',
also depicted in this figure, represents the constraint on
($M_G,c$) from EW oblique corrections
\cite{Davoudiasl:2000jd,Han:2000gp}, but is
subject to assumptions on the high-energy structure of the theory.
The region to the left is ``forbidden''.

Finally, the signature space of the considered $Z^\prime$ models
is the (rather narrow) area delimited by the curves labeled
``$Z^\prime_{\rm ALR}$'' and ``$Z^\prime_\psi$''. Actually,
being characterized by coupling constants that are numerically fixed,
all of the $Z^\prime$ models considered here are represented
simply by curves in the ($M_R,N_S$) plane. In order not to
make the figure too busy, we represent just the leftmost and
rightmost $Z^\prime$ lines, the others lie inside that
domain. Tevatron experimental limits on masses lie around
(or below) 1 TeV and are not reported in this figure.

In practice, in Fig.~\ref{fig3-N-14-100}, the
$X=10^{-5}-10^{-1}$ sneutrino domain is so wide as to
fully include the $Z^\prime$ signature space and, in turn, the
latter fully includes the RS resonance signature space if the
restriction from the $\Lambda_\pi$ line is strictly applied.
From Fig.~\ref{fig3-N-14-100} one can easily read off the domain
in ($M_{\tilde\nu},X$) where, for example, the RS graviton
exchange within its allowed parameter range can mimic sneutrino
exchange by the same number of signal events, as well as the
domain where sneutrino exchange can similarly be confused with
that of a $Z^\prime$. The intersection domain of the three
signature spaces
is the confusion region where the number of events,
$N_S(pp\to\tilde\nu  \to l^+l^-)=N_S(pp\to Z^\prime \to l^+l^-)
=N_S(pp\to G\to l^+l^-)$ ($l=e,\mu$ combined), and the three
spin hypotheses cannot be distinguished from each other
from event rates only. Of course, outside the
confusion regions, the various scenarios {\it may} be
mutually distinguished from one another on the basis of 
the number of signal events $N_S$, and the corresponding domains 
can be readily read off from 
Fig.~\ref{fig3-N-14-100}. For example, events falling to the
right of the line ``$Z^\prime_{\rm ALR}$'' can unambiguously be
identified as sneutrino exchange, and the same is true
for those to the left of the ``$Z^\prime_\psi$'' line
(where we also rely on the $\Lambda_\pi$ constraint);
conversely, events between the ``$c=0.1$'' and
``$Z^\prime_{\rm ALR}$'' lines can be confused as either
sneutrino or $Z^\prime$ events, those between the ``$\Lambda_\pi$''
and ``$c=0.1$'' curves can be either RS graviton or $Z^\prime$ or
sneutrino exchange.

The line labelled ``Excl.\ spin-2'' represents the application
of angular analysis, specifically the $A_{\rm CE}$ condition
(\ref{chisquare}), to the determination of the minimum number
of events needed for the 95\% CL exclusion of the spin-2 RS
graviton hypothesis by the spin-0 sneutrino assumption for
the observed peak at $M=M_R$ (here, the condition
$\Lambda_\pi<10$ TeV has been relaxed). A larger (or equal)
number of events than represented by that curve is therefore
needed to that purpose. Analogously, the curve labelled as
``Excl.\ spin-1'' corresponds to the minimal
number of events necessary to exclude the spin-1 $Z^\prime$
hypothesis once the sneutrino spin-0 scenario has been assumed as
true. For $N_S$ below these curves, confusion regions between 
the models still remain, which are marked in the figure. 

Accordingly, Fig.~\ref{fig3-N-14-100} indicates that
the identification of the sneutrino against the RS graviton 
by angular analysis requires for $M_{\tilde\nu}$ in the range 
from about 1.6 to 3.2 TeV, a minimum of 150-200 events 
(200 events if one strictly applies the condition 
$\Lambda_\pi<10$ TeV).
The exclusion of the considered $Z^\prime$ models in the 
confusion dilepton mass range can be
obtained with a minimum of about 120 events. The full
identification of a sneutrino can finally be
determined by the exclusion of both, namely, by the larger
of these numbers. The complete sneutrino identification domain 
in the ($M_{\tilde\nu},X$) plane at
$\sqrt s=14$ TeV and ${\cal L}_{\rm int}=100\,\text{fb}^{-1}$
will later be presented, and compared with the case of the same LHC
energy, but lower luminosity.

In this analysis, we have adopted a mass-independent $K$-factor. 
This is clearly a simplification. In fact, it has been shown that the $K$-factor is larger than 1.3 at low masses, but then falls with increasing sneutrino mass \cite{Choudhury:2002aua,Dreiner:2006sv,ShaoMing:2006dx,Chen:2006ep}. 
For the considered asymmetry, $A_\text{CE}$, this effect cancels out, but it does affect the overall number of events, and hence the discovery limit, as discussed above.
\section{Lower luminosity and energy}
\label{sec:low-energy}
\setcounter{equation}{0}
\subsection{LHC nominal energy and lower luminosity}
The panels in the left column of
Fig.~\ref{fig4-N-7-14-10} represent the
signature spaces for sneutrino, $Z^\prime$ and RS graviton
exchanges at the LHC with
$\sqrt s=14$ TeV and ${\cal L}_{\rm int}=10\,\text{fb}^{-1}$.
The same experimental cuts as before, and $l=e,\mu$ channel
combination, have been applied. The curves
enclosing the respective domains are labelled quite similar to
Fig.~\ref{fig3-N-14-100}, and in particular: the upper panel shows
the confusion region between the sneutrino and spin-2 RS hypotheses
and the dashed ``Excl.\ spin-2'' line gives the minimum number of events
required for distinguishing them from each other by the $A_{\rm CE}$
angular analysis; analogously, the middle panel shows the
$\tilde\nu$--$Z^\prime$s confusion region and the dashed ``Excl.\ spin-1''
line is the minimum number of events for a discrimination
among them; finally, the lower panel shows the area of total
confusion, where all three hypotheses give the same number of
events under the peak at $M_R$. Clearly, compared to the previous
case of high luminosity, the sensitivity to sneutrino and its
identification via $A_{\rm CE}$ occurs for lower 
$M_{\tilde\nu}$, about 1.9--2.7 TeV, while the minimal number 
of needed events remains practically the same.

\begin{figure}[htbp!] 
\centerline{ 
\hspace*{2.0cm}
\includegraphics[width=7.cm,angle=0]{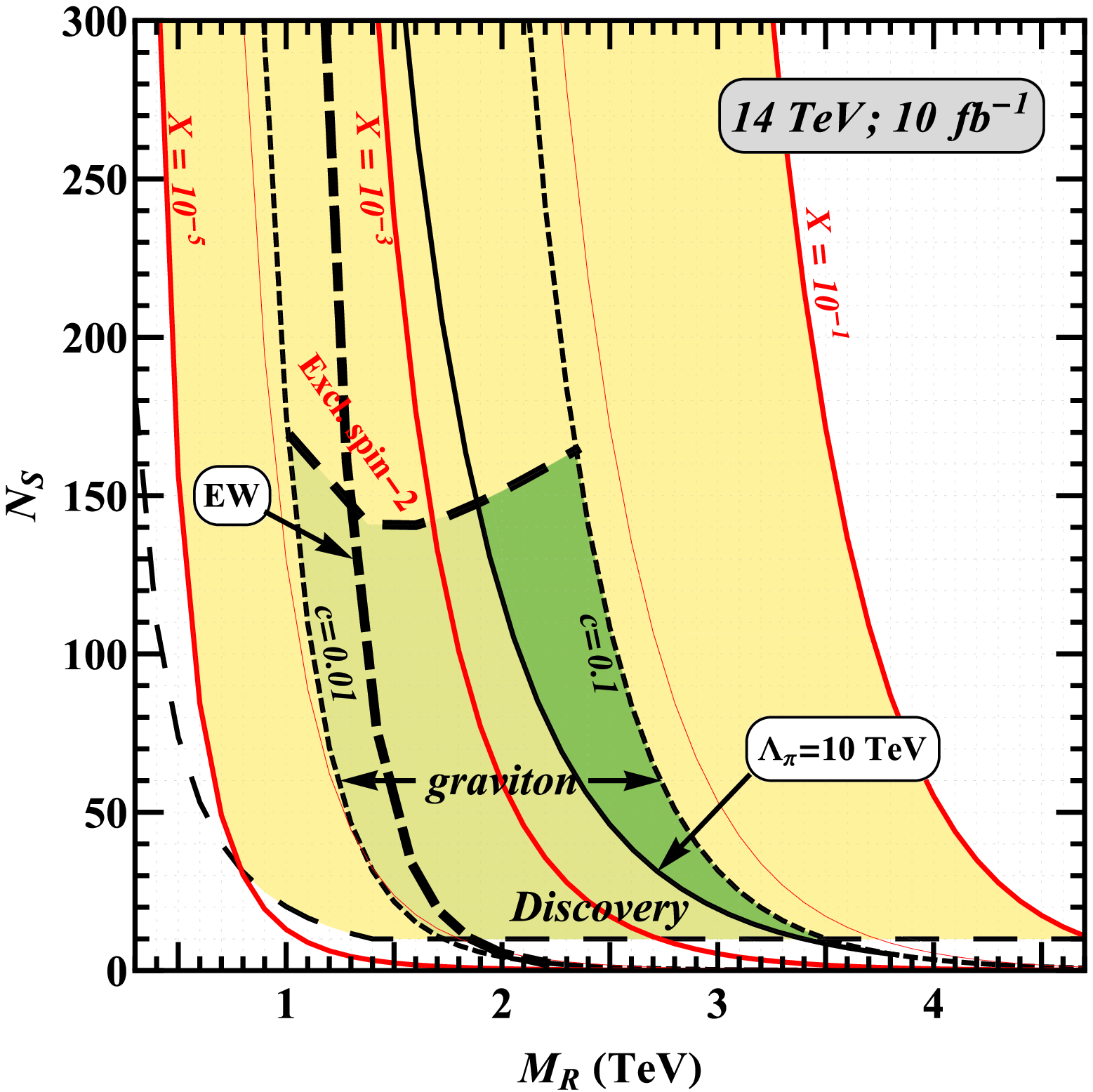}
\hspace*{0.4cm}
\includegraphics[width=7.cm,angle=0]{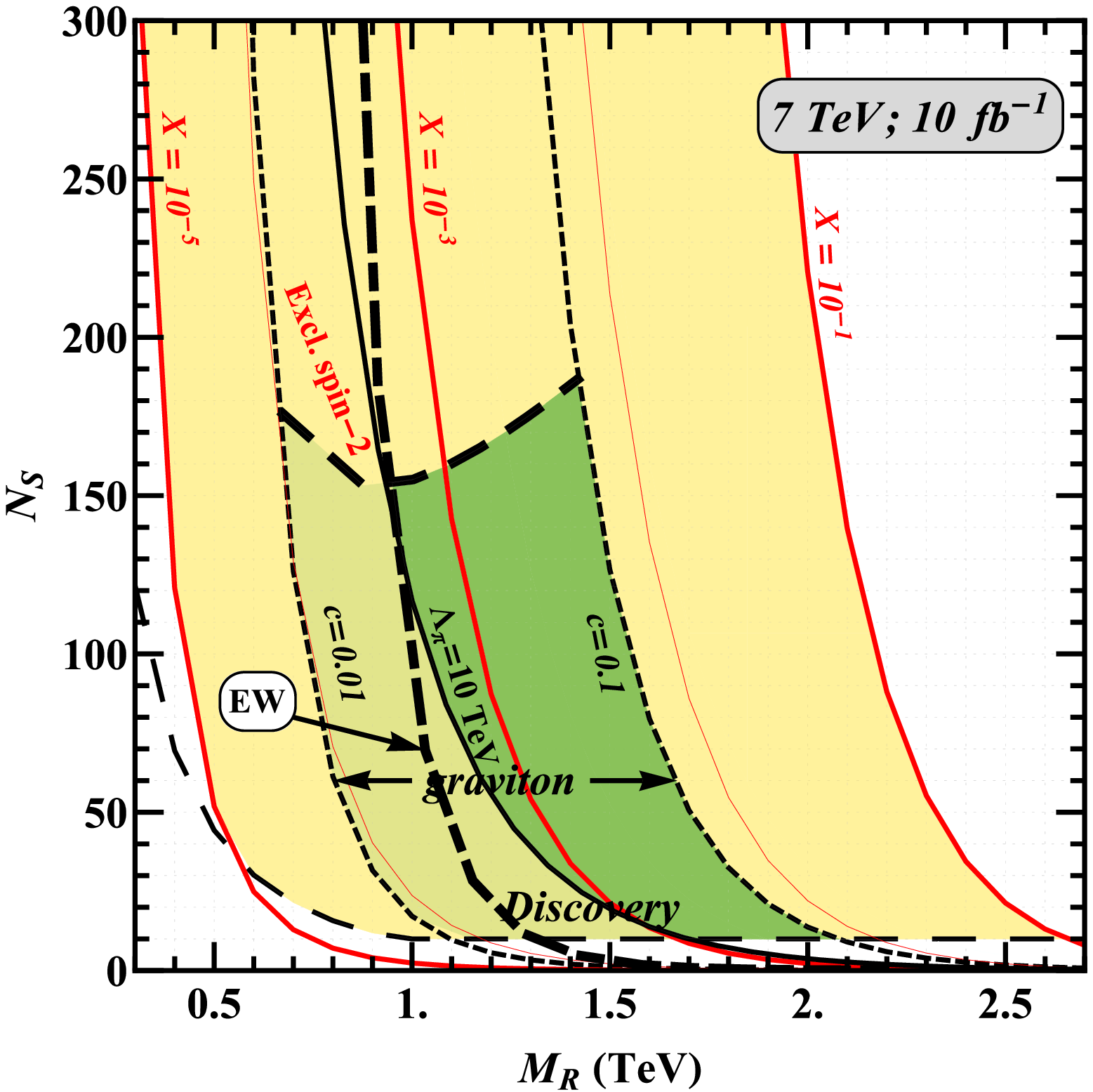}}
\centerline{ 
\hspace*{2.0cm}
\includegraphics[width=7.cm,angle=0]{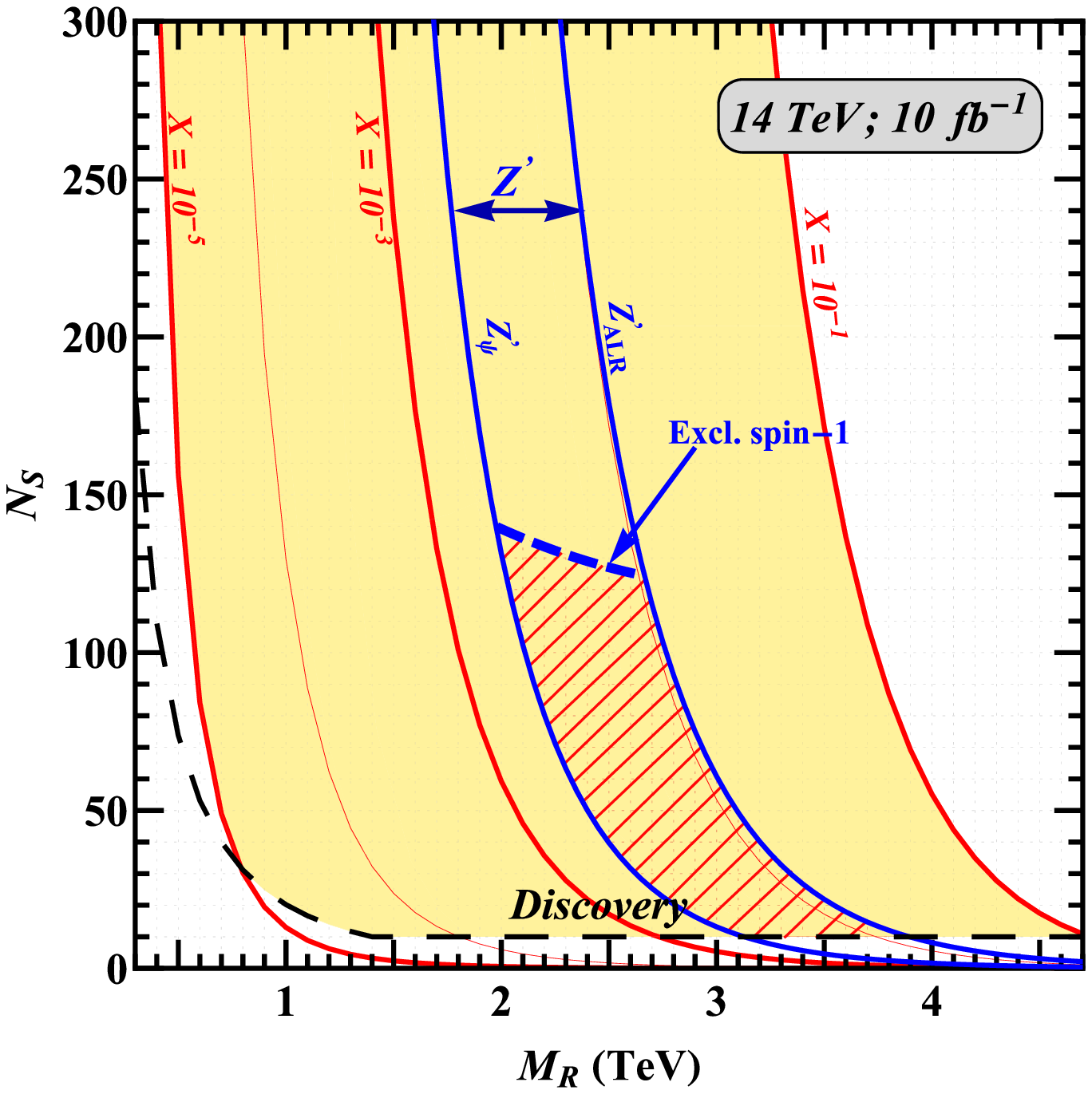}
\hspace*{0.4cm}
\includegraphics[width=7.cm,angle=0]{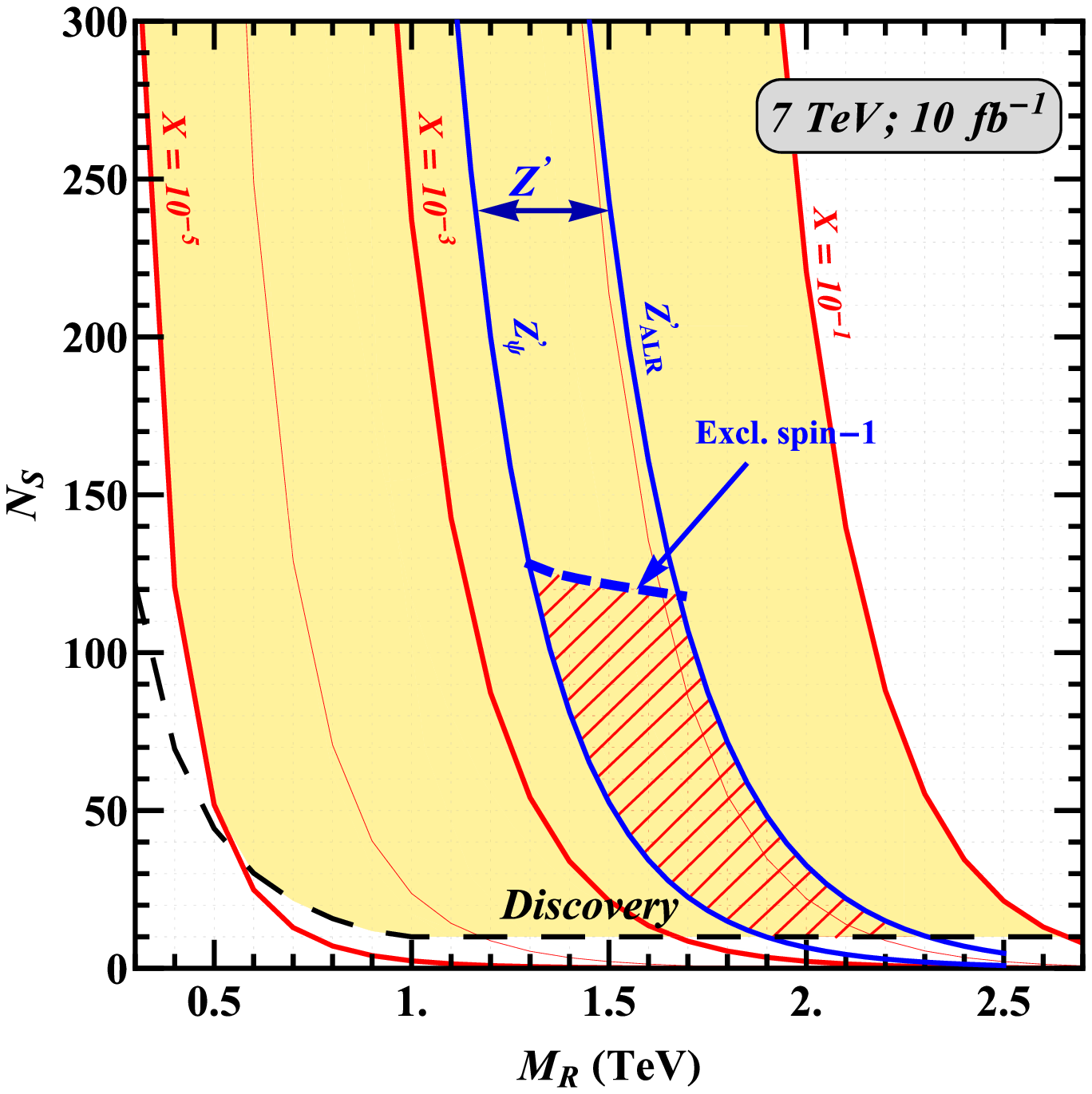}}
\centerline{ 
\hspace*{2.0cm}
\includegraphics[width=7.cm,angle=0]{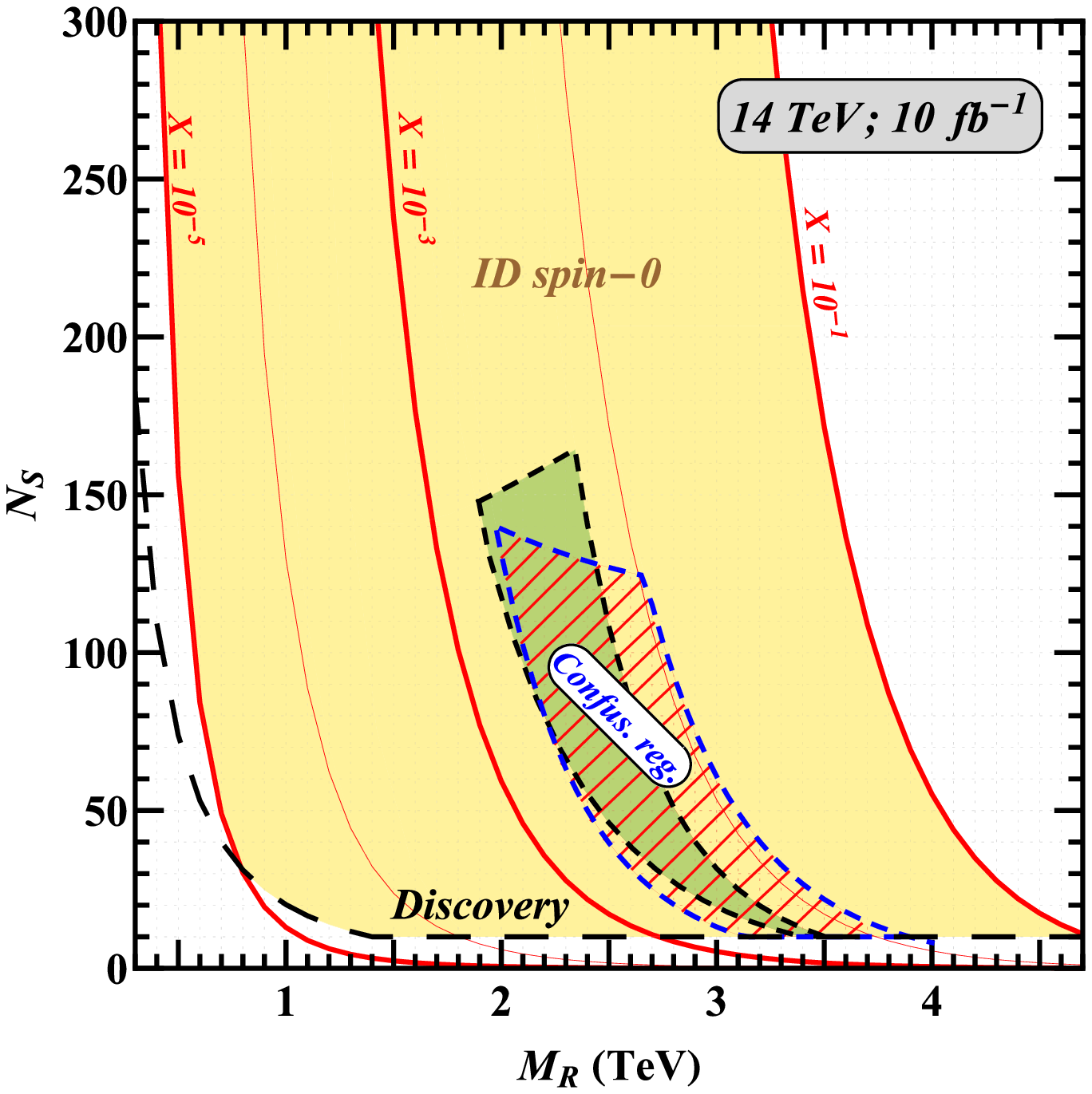}
\hspace*{0.4cm}
\includegraphics[width=7.cm,angle=0]{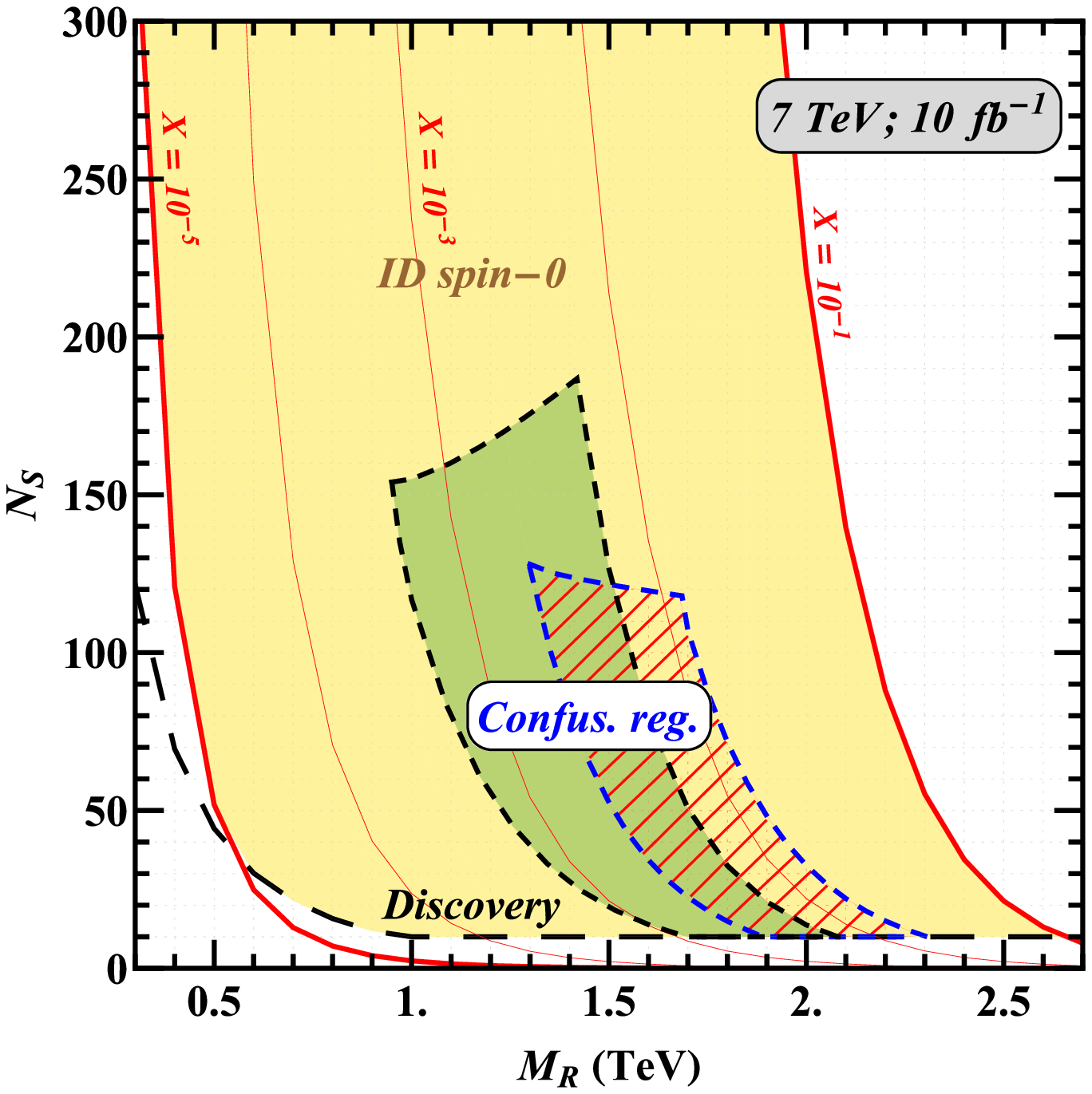}}
\caption{\label{fig4-N-7-14-10}
Same as in Fig.~\ref{fig3-N-14-100} but for $\tilde\nu_\tau$ vs.\
$G$ (top panels), $\tilde\nu_\tau$ vs.\ $Z^\prime$ (middle panels)
and both combined (bottom panels) at the LHC with
$\Lumint=10~\text{fb}^{-1}$, $\sqrt{s}=14$ TeV (left column) and
$\sqrt{s}=7$ TeV (right column). Note the different mass scales.
}
\end{figure}

\begin{figure}[h!] 
\centerline{ \hspace*{-0.0cm}
\includegraphics[width=7.cm,angle=0]{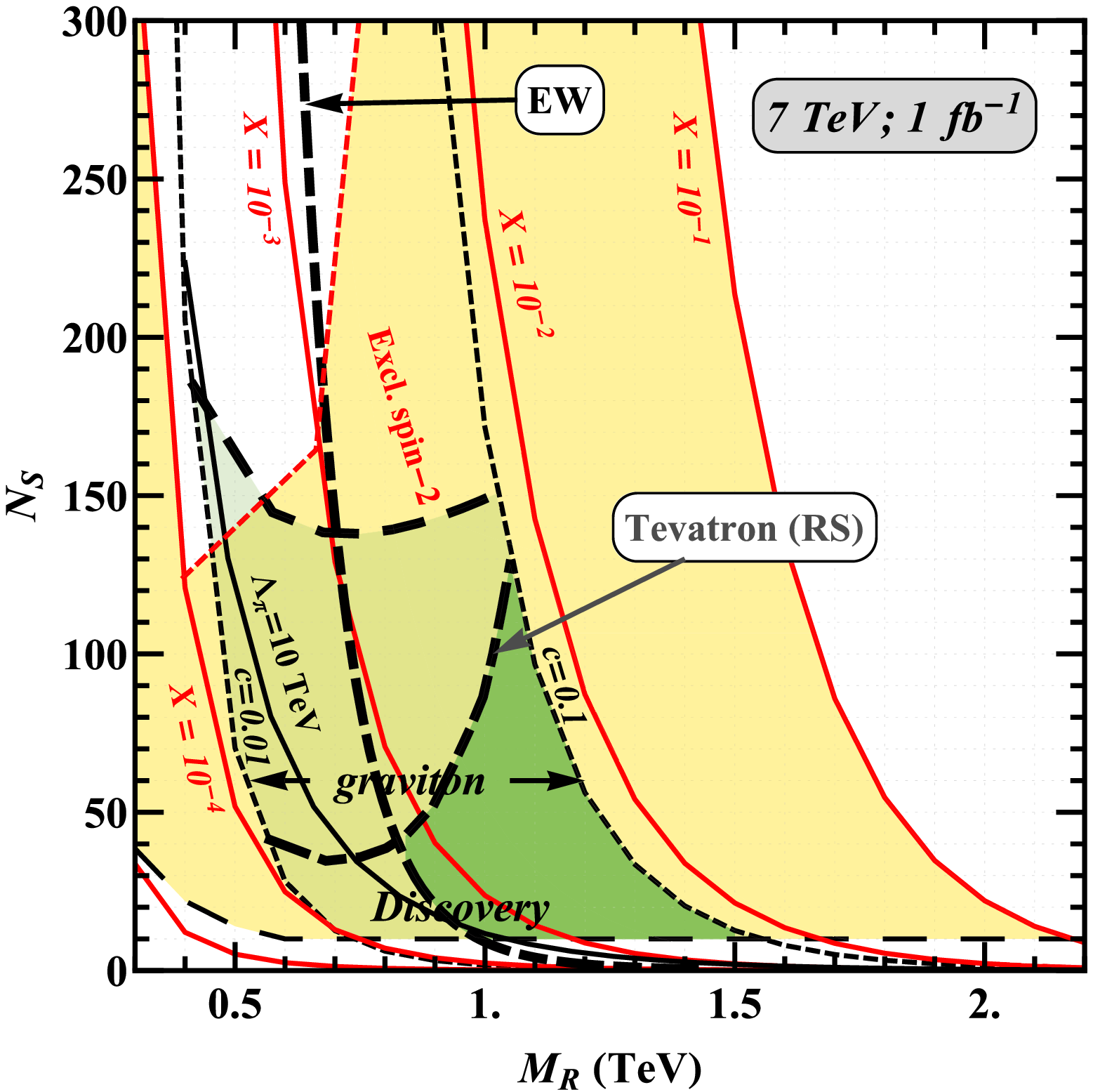}}
\centerline{\includegraphics[width=7.cm,angle=0]{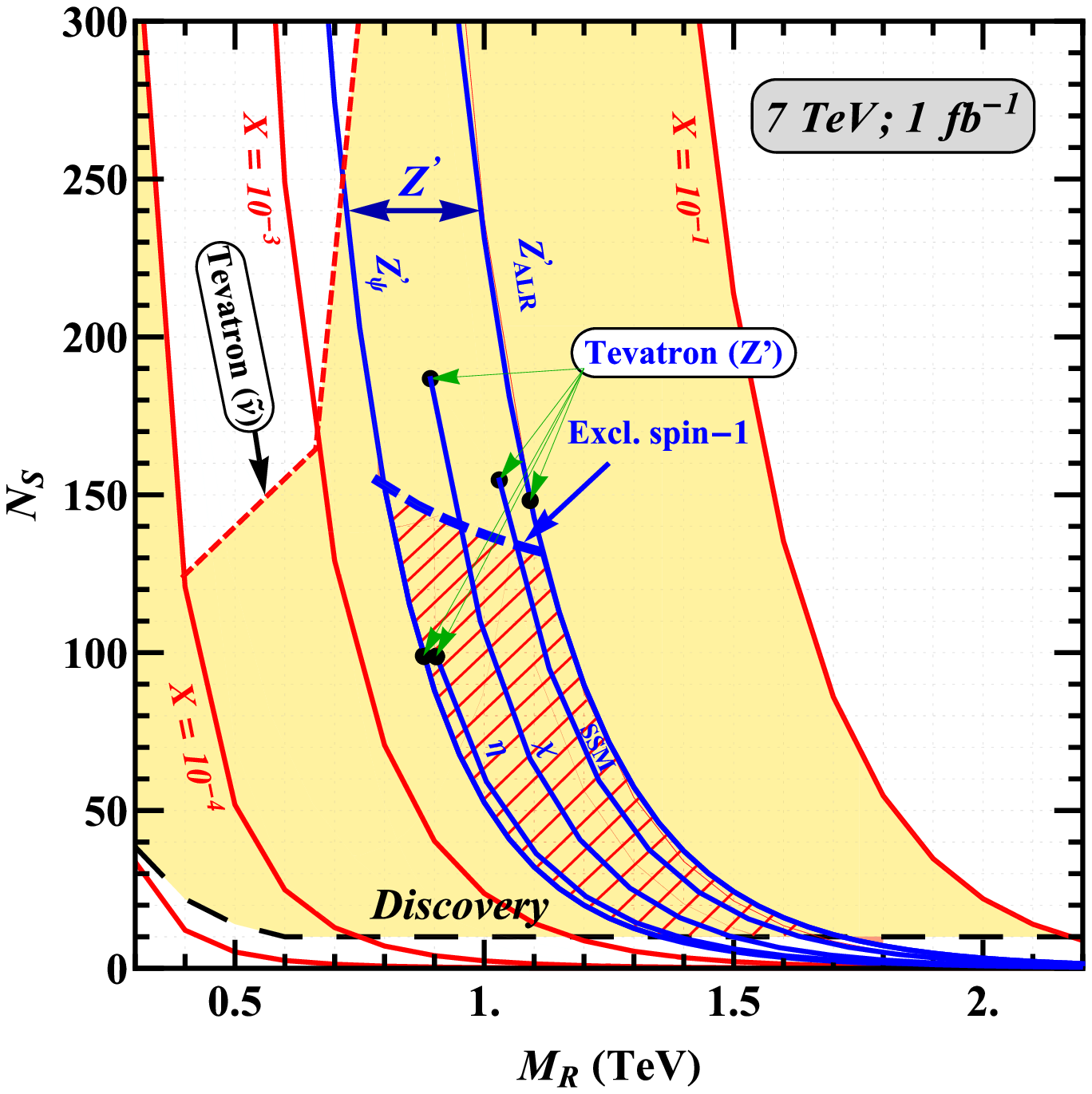}}
\centerline{\includegraphics[width=7.cm,angle=0]{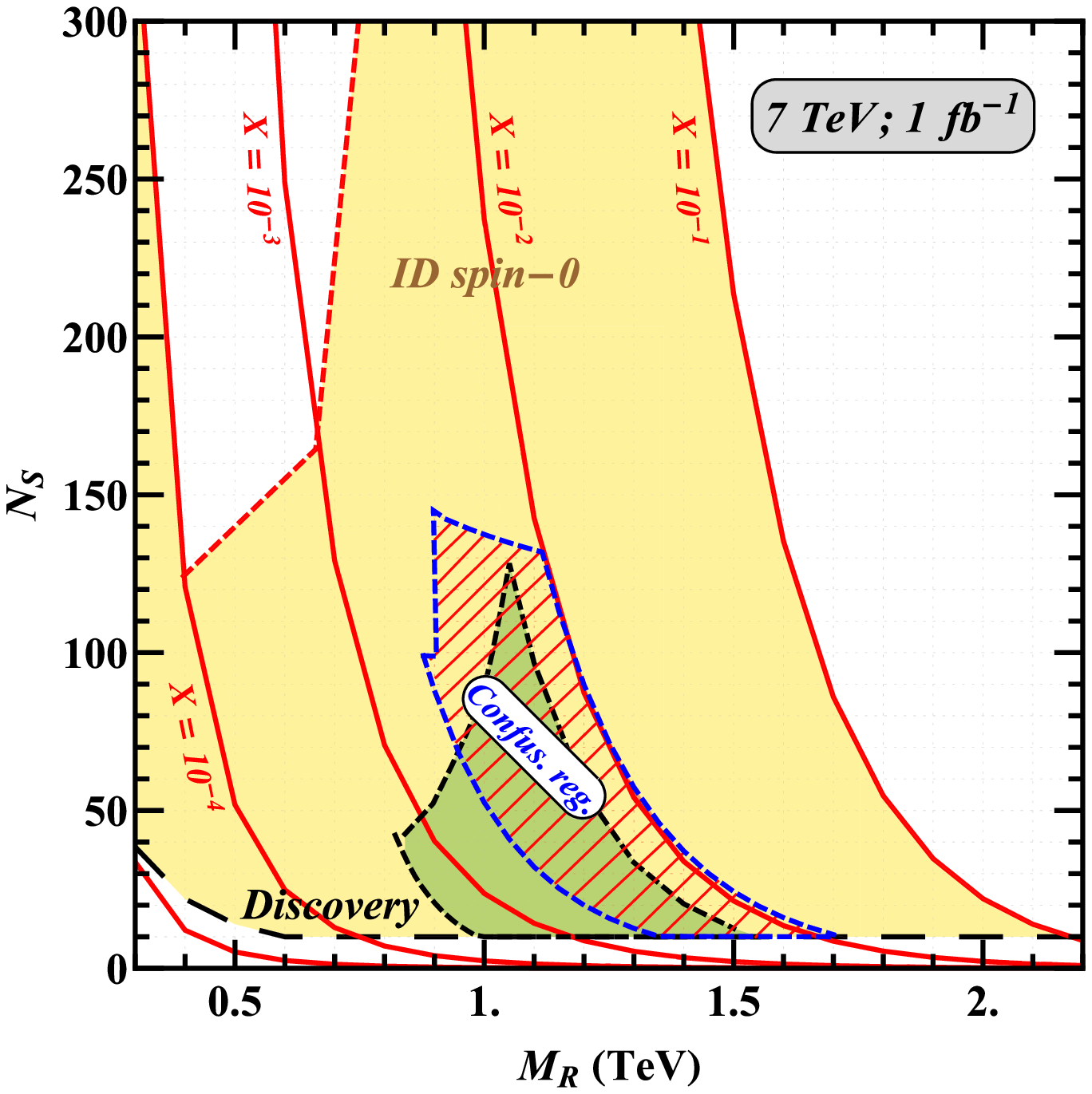}}
\caption{\label{fig6-N-7-1} Same as in Fig.~\ref{fig4-N-7-14-10}
but for the LHC with $\sqrt{s}=7$ TeV and
$\Lumint=1~\text{fb}^{-1}$.
Constraints on the signature spaces from the current experimental
data are also shown.}
\end{figure}

\subsection{Lower energy and lower luminosity}
In Fig.~\ref{fig4-N-7-14-10}, the  panels in the right column 
represent in complete analogy with those on the left the 
signature spaces and confusion regions for the initially available 
LHC energy, $\sqrt s=7$ TeV, and luminosity 
${\cal L}_{\rm int}=10\,\text{fb}^{-1}$. The same phase space 
cuts (and $e$, $\mu$ channel combination) have been applied as in 
the preceding sections, and the same notations are used to label 
the curves delimiting the respective domains. The only difference 
from the panels on the left discussed above, is that, in
the case of the RS graviton, the ``EW'' constraining curve has
considerably approached the ``$\Lambda_\pi$'' one.
We recall that it is the area to the left or below, which is
disfavored.

The same kind of conclusions regarding the pairwise 
confusion regions between the three hypotheses,
the overall confusion region and the sensitivity for spin-0
sneutrino identification can be derived. The
minimal numbers of events needed for exclusion of the 
$Z^\prime$ and RS graviton hypotheses by angular analysis 
are comparable to the
previous cases of higher LHC energy, but the relevant values of
$M_{\tilde\nu}$ are now constrained in the range 1.0--1.7 TeV, 
approximately, close to (but still above) the current Fermilab 
Tevatron discovery limits.

It may take some time before the experiments will be able to
collect the desirable integrated luminosities of data considered 
above. In Fig.~\ref{fig6-N-7-1}, we plot the signature spaces 
and confusion regions for $\sqrt s=7$ TeV and the still lower 
luminosity ${\cal L}_{\rm int}=1\,\text{fb}^{-1}$, under 
the same assumptions for the foreseeable experimental cuts 
on the phase space. The analysis is now complicated by the fact 
that the relevant masses $M_R$ enter the range of the current 
Tevatron limits and these, therefore, as shown by the figure, 
have a significant role in constraining the
allowed areas. Basically, one can see that the minimal number of
events needed to distinguish the sneutrino from the RS graviton
resonance is well above the limit allowed by the Tevatron, 
see the upper panel. An analogous situation is met for 
exclusion of some of the $Z^\prime$ models, which are in 
the middle panel represented in detail by their respective 
characteristic lines, stopped at the Tevatron lower limits. 
The application of the angular analysis for
spin discrimination in the confusion regions is, accordingly, 
somewhat problematic in this low-energy and low-luminosity 
case.

Indeed, the dashed ``Excl.\ spin-2'' line in the upper panel in 
Fig.~\ref{fig6-N-7-1} shows that a minimum of 140 dilepton events 
should be necessary to discriminate spin-0 from spin-2 by the 
$A_{\rm CE}$ angular analysis, but this is well above the limits 
in $N_S$ for the relevant values of $M=M_R$, as set by the 
curve ``Tevatron(RS)''. Regarding the discrimination from 
the $Z^\prime$ hypothesis, the figure in the middle panel, that 
includes also the current Tevatron limits on the sneutrino mass, 
shows that for selected values of $M_R$ the spin-0 assumption may 
indeed be discriminated from some of the $Z^\prime$ models {\it via} 
angular analysis, for $N_S$ higher than the ``Tevatron($Z^\prime$)'' 
line. Taking into account the above restrictions 
from the $A_{\rm CE}$-angular analysis, the sneutrino 
identification domain in this case is the portion of the 
sneutrino signature space depicted in yellow in the bottom panel 
of Fig.~\ref{fig6-N-7-1}.

\begin{figure}[htb!]
\centerline{
\includegraphics[width=10cm,angle=0]{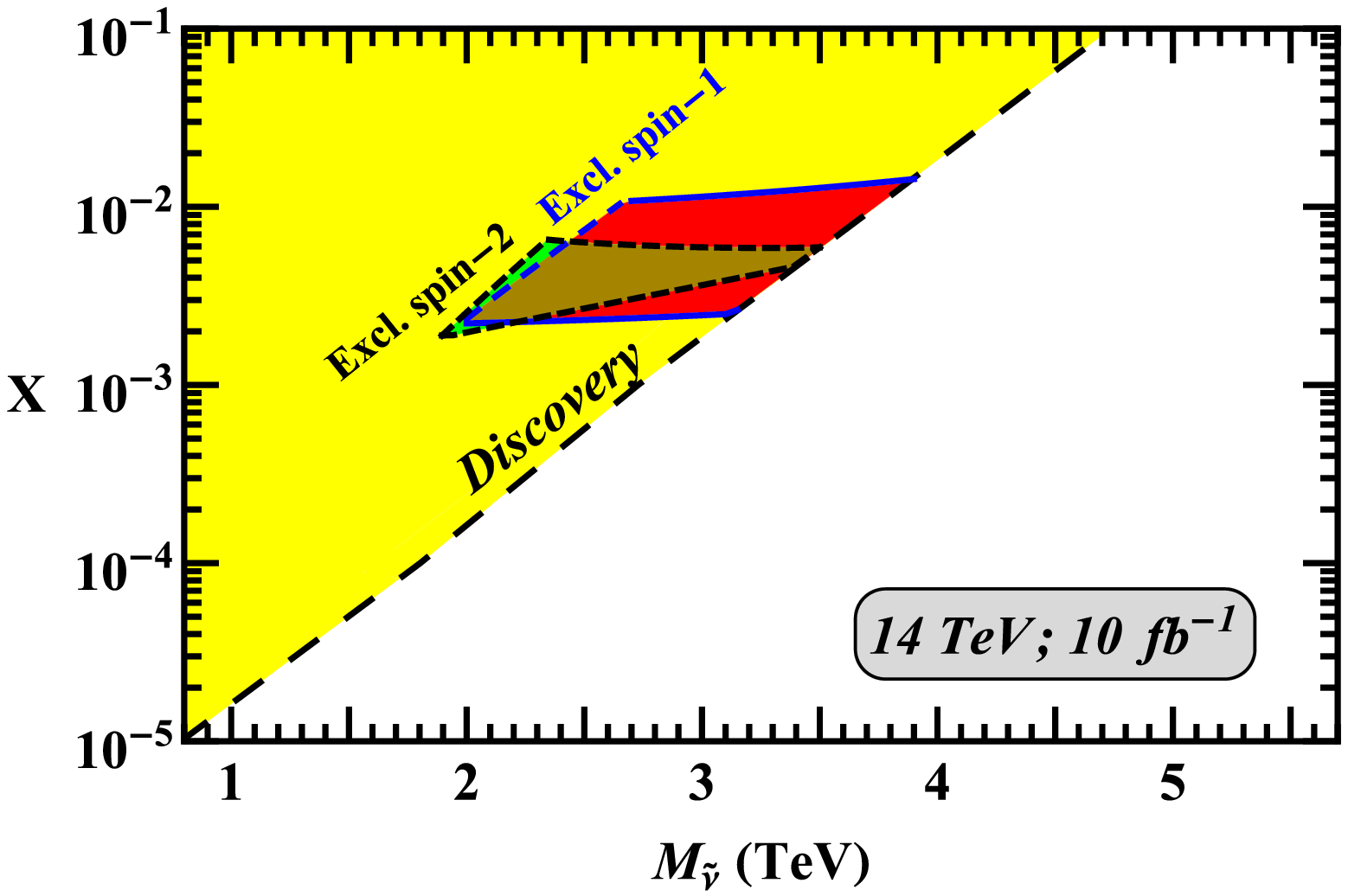}}
\vspace*{0.1cm}
\centerline{\includegraphics[width=10cm,angle=0]{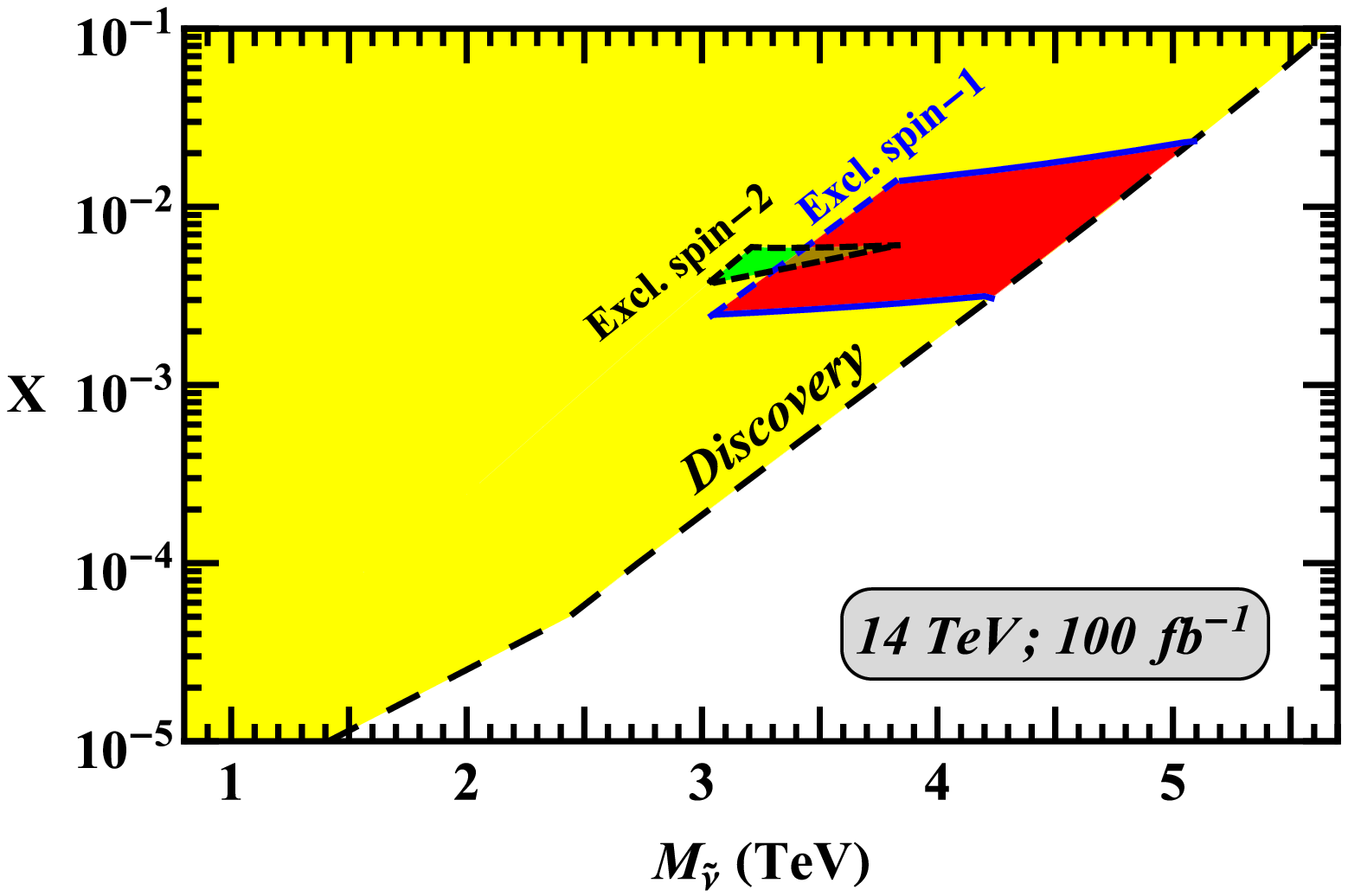}
}
\caption{\label{fig5-X-Mnu} Discovery reach (long-dashed line) 
and spin-0 sneutrino identification domain (yellow) in the 
($M_{\tilde\nu}$,$X$) plane obtained from lepton pair production 
($l = e, \mu$) at the LHC with $\sqrt{s}=14$ TeV, $\Lumint=10$ fb$^{-1}$
(top panel) and $\Lumint=100$ fb$^{-1}$ (bottom panel), using the
production cross section and center--edge asymmetry. The discovery
limit is defined by $5\sqrt {N_{SM}}$ or by 10 events, whichever
is larger.}
\end{figure}

\section{Luminosity and coupling strength issues}
\label{Sec:coupling}

For an assessment of the role of luminosity, we show in the two 
panels of Fig.~\ref{fig5-X-Mnu} the sneutrino discovery and 
identification reaches in the ($M_{\tilde\nu},X$) plane for 
the nominal energy $\sqrt s=14$ TeV and two values of integrated luminosity,
${\cal L}_{\rm int}=10,\,{\rm and}\,100~\text{fb}^{-1}$. 
The differently colored areas essentially reflect the comments 
made with regard to Figs.~\ref{fig3-N-14-100} and 
\ref{fig4-N-7-14-10}, concerning the confusion and spin-2 and spin-1 
exclusion areas determined there. 

Specifically, the yellow domain represents the values 
of $M_{\tilde\nu}$ and $X$ for which, if observed as a peak 
in the process (\ref{proc_DY}), the spin-0 sneutrino can 
unambiguously be identified: referring to the example of 
Fig.~\ref{fig3-N-14-100}, it is composed of the subdomain 
where identification can be obtained simply from the event rates 
by themselves, and that where the $A_{\rm CE}$-based 
angular analysis is needed to perform the discrimination 
in the confusion regions with the two competitor hypotheses. 
The trapezoidal and triangular areas (red, brown 
and green) represent the 
domains in the sneutrino parameter space where 
the distinction from both the $Z^\prime$ and the 
RS graviton hypotheses, cannot be done. 

\begin{figure}[htb!] 
\centerline{ 
\hspace*{-0.5cm}
\includegraphics[width=10.0cm,angle=0]{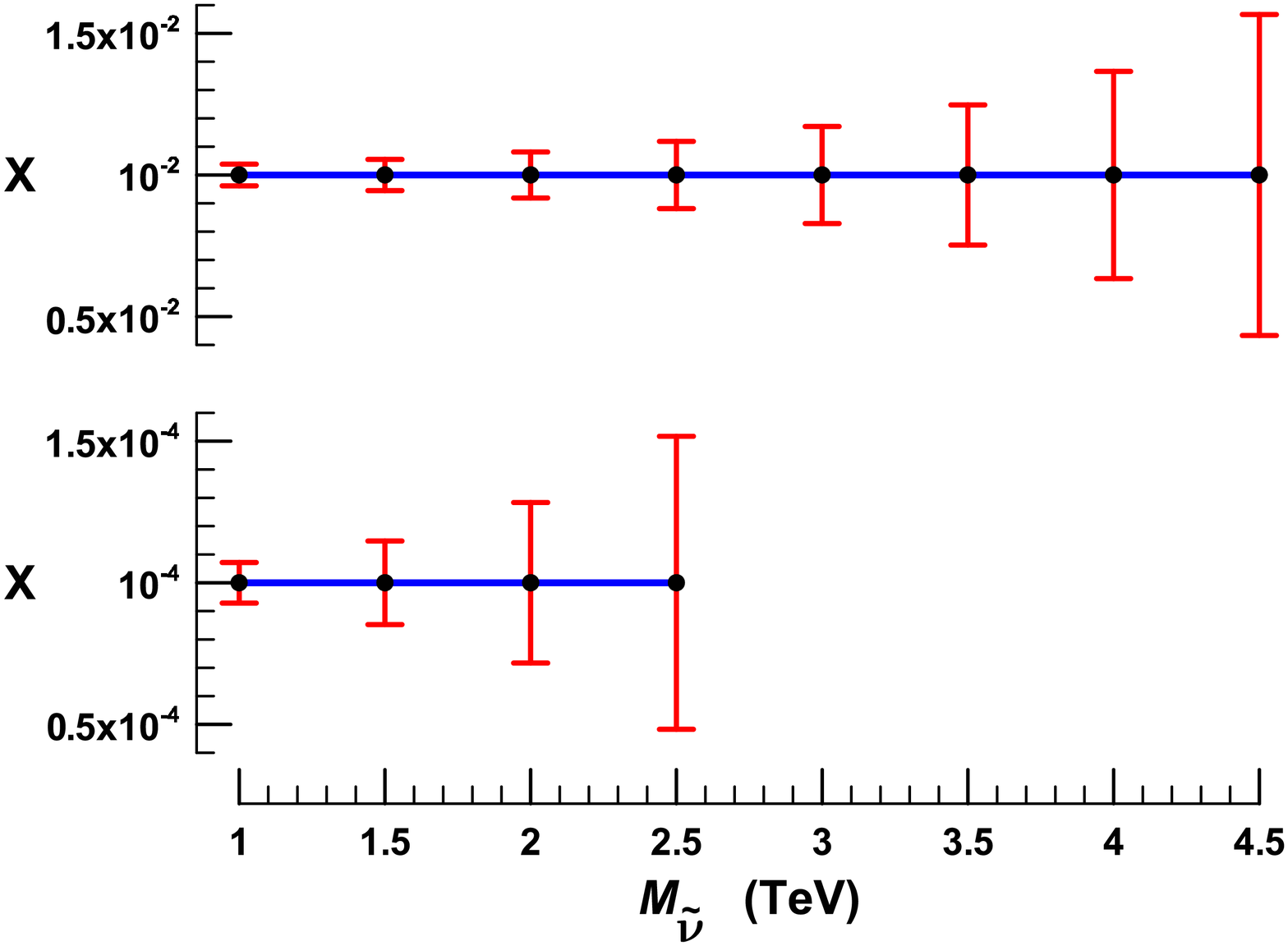}}
\caption{\label{precision} Precision of the spin-0 sneutrino
model parameter $X$ for $X=10^{-2}$ and $X=10^{-4}$ for increasing
${\tilde\nu}$ mass obtained at the LHC with $\sqrt{s}=14$ TeV and
$\Lumint=100$ fb$^{-1}$.}
\end{figure}
Assuming that the data is indeed consistent
with the spin-0 sneutrino prediction for certain values of $M_{\tilde\nu}$ and $X$, one
would also like to know how precisely the parameter $X$  of Eq.~(\ref{X}) can be
constrained. This question is addressed in Fig.~\ref{precision},
for $\sqrt{s}=14$ TeV and $\Lumint=100$ fb$^{-1}$, and
for two values of $X$: $10^{-4}$ and
$10^{-2}$. The range of sneutrino masses considered, is obviously
within the discovery reach, cf.\ Fig.~\ref{fig0}. As the
sneutrino mass is increased, the number of events decreases
drastically and the statistical fluctuations increase,
becoming as large as $\vert\delta X\vert/X\approx 0.5$ at the highest masses shown.

In summary, we have determined the regions of mass and 
coupling strengths for which a sneutrino, if produced 
resonantly at the LHC, could be distinguished from other 
candidate resonances such as $Z^\prime$s and Randall--Sundrum 
excited gravitons. Two energies and three values of the 
integrated luminosity have been considered. Of the order of 
200 events are required for a discrimination against both 
alternative interpretations in the regions of confusion 
via the $A_{\rm CE}$-based angular analysis. At the nominal 
energy of 14~TeV, and with $100~\text{fb}^{-1}$, this 
corresponds to the mass range about 3.0--3.8~TeV,
and 1.9--2.6~TeV at $10~\text{fb}^{-1}$. Outside these mass 
ranges sneutrino identification can be determined by event rates 
only. 

\goodbreak
\vspace{0.5cm}
\leftline{\bf Acknowledgements}
\par\noindent
A useful correspondence with Chris Hays is gratefully acknowledged.
This research has been partially supported by the Abdus Salam
ICTP and the Belarusian Republican Foundation for Fundamental
Research. The work of PO has been supported by The Research Council of
Norway, and that of NP by funds of MiUR and of University of Trieste.


\end{document}